\documentclass{aa}
\usepackage[varg]{txfonts}
\usepackage{natbib, amsmath, amssymb, amsfonts, graphicx}
\usepackage{mathtools}
\usepackage{subfig}
\usepackage{comment}
\usepackage{xcolor}
\usepackage{diagbox} 
\usepackage[citecolor=blue, linkcolor=blue, urlcolor = black, colorlinks = true, backref = page]{hyperref}
\usepackage{bigints}
\usepackage{lscape}
\usepackage{fancyvrb}
\usepackage{siunitx}

\RecustomVerbatimCommand{\VerbatimInput}{VerbatimInput}%
{fontsize=\footnotesize,
 frame=lines,  
 framesep=2em, 
 rulecolor=\color{gray},
 commandchars=\|\(\), 
 commentchar=*        
}

\title{Hydrodynamic modelling of dynamical tides dissipation in Jupiter's interior as revealed by Juno}
\author{H. Dhouib \inst{1}
\and C. Baruteau  \inst{2}
\and S. Mathis    \inst{3}
\and F. Debras    \inst{2}
\and A. Astoul    \inst{4}
\and M. Rieutord  \inst{2}
}

\institute{Université Paris Cité, Université Paris-Saclay, CEA, CNRS, AIM,  F-91191, Gif-sur-Yvette, France\\\email{hachem.dhouib@cea.fr}
\and IRAP, Université de Toulouse, CNRS UMR 5277, UPS, F-31400 Toulouse, France
\and Université Paris-Saclay, Université
Paris Cité, CEA, CNRS, AIM,  F-91191, Gif-sur-Yvette, France
\and Department of Applied Mathematics, School of Mathematics, University of Leeds, Leeds LS2 9JT, UK}

\titlerunning{Hydrodynamic modelling of dynamical tides dissipation in Jupiter's interior as revealed by Juno}
\authorrunning{H. Dhouib et al.}

\abstract
{The Juno spacecraft has acquired exceptionally precise data on Jupiter's gravity field, offering invaluable insights into Jupiter's tidal response, interior structure, and dynamics, establishing crucial constraints.}
{We develop a new model for calculating Jupiter's tidal response based on its latest interior model, while also examining the significance of different dissipation processes for the evolution of its system. We study the dissipation of dynamical tides in Jupiter by thermal, viscous and molecular diffusivities acting on gravito-inertial waves in stably stratified  zones and inertial waves in convection ones.}
{We solve the linearised equations for the equilibrium tide. Next, we compute the dynamical tides using linear hydrodynamical simulations based on a spectral method. The Coriolis force is fully taken into account, but the centrifugal effect is neglected. We study the dynamical tides occurring in Jupiter using internal structure models that respect Juno's constraints. We study specifically the dominant quadrupolar tidal components and our focus is on the frequency range that corresponds to the tidal frequencies associated with Jupiter's Galilean satellites.}
{By incorporating the different dissipation mechanisms, we calculate the total dissipation and determine the imaginary part of the tidal Love number. We find a significant frequency dependence in dissipation spectra, indicating a strong relationship between dissipation and forcing frequency. Furthermore, our analysis reveals that, in the chosen parameter regime in which kinematic viscosity, thermal and molecular diffusivities are equal, the dominant mechanism contributing to dissipation is viscosity, exceeding in magnitude both thermal and chemical dissipation. We find that the presence of stably stratified zones plays an important role in explaining the high dissipation observed in Jupiter.}
{}

\begin{document}
\maketitle

\section{Introduction}
Tidal interactions between Jupiter and its Galilean satellites are recognised as influential factors in both the system's orbital evolution and the internal dynamics \citep[e.g.][]{Lainey2009}.
Traditionally, the tidal response of a gaseous star or planet like Jupiter is treated using the concept of equilibrium tide \citep{Zahn1966, Zahn1989, Remus2012}, where hydrostatic deformation exhibits a minor phase lag in response to the dissipative processes caused by tidal forcing. However, the observed strong tidal dissipation in Jupiter \citep{Lainey2009} and the gravitational perturbations recently measured by the Juno spacecraft \citep{Durante2020} cannot be fully explained by the equilibrium tide alone.
In fact, the Juno spacecraft has not only enhanced our understanding of Jupiter's tidal dynamics but has also allowed us to delve deeper into the gravitational perturbation and tidal dissipation phenomena associated with the planet.
On the one hand, it has acquired precise measurements of Jupiter's tidal Love numbers, $k_{\ell m}$, which quantitatively characterize the planet's response to tidal forcing represented by spherical harmonics of degree $\ell$ and order $m$.
By analysing the real part of these Love numbers, we gain valuable information about the gravitational perturbations experienced by Jupiter. On the other hand, the imaginary part of the Love numbers provides us with insights into the processes of tidal dissipation occurring within the planet.
Recently, \cite{Durante2020} measured the Love number value for the dominant tidal component $k_{22} = 0.565\pm 0.018$ ($3\sigma$ uncertainty). This is lower than the theoretical hydrostatic value of $k_{22}^{(\rm eq)} = 0.589$ as stated by \cite{Wahl2020}, indicating a difference of approximately $\Delta k_{22}\approx-4\%$. \cite{Wahl2020} noted that the influence of the interior structure on $k_{\ell m}$ is negligible when considering models that accurately reproduce the zonal harmonics $J_2$, $J_4$, and $J_6$, which have already been measured with high precision by Juno. 
This discrepancy between the observed and the computed values of $k_{22}$ could potentially be attributed to the influence of dynamical tides \citep{Zahn1975, Ogilvie2004}. Indeed, the conventional concept of the equilibrium tide, does not satisfy the full equation of motion because the acceleration of the fluid is neglected \citep{Zahn1966b}. Hence, a comprehensive understanding of the planet's tidal response requires the inclusion of corrections. These corrections introduce wavelike motions within the planet and depend on both the tidal frequency and the internal structure \citep{Ogilvie2014}. The dynamical (wavelike) tide offers additional channels for tidal dissipation and generates supplementary gravitational perturbations, surpassing the effects solely attributed to the hydrostatic deformation \citep{Idini2021,  Lai2021, Lin2023}.

The detection of gravitational signatures from dynamical tides can provide valuable insights into the interior structure of Jupiter, as it is influenced by both the tidal frequency and the internal structure. 
\citet{Idini2021,  Lai2021, Idini2022b, Idini2022a, Dewberry2022, Dewberry2023, Lin2023} have found that $\Delta k_{22}$ can be largely attributed to the Coriolis effect on the fundamental modes (f-modes). Additionally, \cite{Idini2022b} have proposed that resonant locking with a gravity mode in an extended diluted core could explain a $\Delta k_{42} \approx -11\%$ difference between the observed and computed values of $k_{42}$. 
This finding supports the existence of a diluted core in Jupiter, a possibility also suggested by \cite{Stevenson1985} and by Juno's measurements of gravitational moments \citep{Wahl2017, Militzer2022}.
The Coriolis force plays a crucial role in Jupiter's tidal responses because tidal frequencies of its Galilean satellites are comparable with the planet's spin frequency. Including the Coriolis force introduces inertial waves in (magnetised) convective regions \citep{Rieutord1997, Ogilvie2004, Ogilvie2009, Ogilvie2013, Rieutord2010, Baruteau2013, Guenel2016a, Guenel2016b, Mathis2016, Wei2016, Wei2018, Lin2018, Astoul2019}  and a combination of gravity waves and inertial waves, known as gravito-inertial waves, in stably stratified regions \citep{Dintrans1999, Dintrans2000, Mathis2009, Auclair2015, Andre2019, Pontin2023}.

The dissipation of dynamical tides occurs through various friction mechanisms, including turbulent friction in convective layers and heat diffusion in stably stratified regions \citep[e.g.][]{Ogilvie2014, Mathis2019, Duguid2020, Vidal2020, Vries2023}. The rate of tidal dissipation in convective and stably stratified regions of planets has significant implications for the evolution of planet-moon systems.
In the context of Jupiter and Saturn systems, our understanding of tidal evolution has undergone a remarkable transformation. Both planets exhibit tidal dissipation that is one or several orders of magnitude stronger than previous predictions based on moon formation scenarios \citep{Goldreich1966}. This intense dissipation is essential to explain their rapid orbital migration, a phenomenon that came to light through precise astrometric measurements \citep{Lainey2009, Lainey2012, Lainey2017, Lainey2020}.
For instance, \cite{Lainey2009} have fitted a dynamical model, including parameterised tidal dissipation, to astrometric observations from 1891 to 2007
of the Galilean satellites. They found that the tidal dissipation is $k_{2 2}/Q = \left(1.1 \pm 0.2 \right)\times 10^{-5}$ (where $Q$ is the quality factor which evaluates the ratio
between the maximum energy stored in the tidal distortion and the energy dissipated during an orbital period), for the asynchronous tide due to Io.

Giant gas planets have traditionally been modelled by a three-layer model. This model entails a central rocky/icy core enveloped by a convective layer comprising metallic hydrogen and helium, which is further encompassed by an outer layer consisting of molecular hydrogen and helium \citep{Stevenson1982, Guillot1994}. While this model serves as a reference for gas giant planets, uncertainties persist regarding the specific size of each region and the precise characteristics of the transitions between them. Recent studies have been diverging from the conventional standard model and delving into alternative interior structures. Specifically, \cite{Leconte2012} proposed a model with a gradient of entropy and heavy elements throughout a whole semi-convective (convective regions that are well-mixed and separated by thin interfaces with stable stratification, creating a staircase-like structure in the entropy profile) planet, suppressing the need for a compact core in Jupiter.
\cite{Stevenson1985, Wahl2017, Debras2019} investigated the possibility of incorporating stable stratification into their models that takes the form of a substantial but diffuse core that extends beyond the previously believed convective zone.
Therefore, significant portions of giant planet interiors are expected to exhibit an unstable entropy gradient, which competes with stable composition gradients. This competition can result in the emergence of double-diffusive convection, leading to the formation of semi-convective layers \citep[e.g.][]{Garaud2018}. Although these layers cannot be directly observed, their formation introduces distinct physics compared to traditional adiabatic models. Consequently, they have a profound impact on the behaviour and subsequent evolution of the system \citep[e.g.][]{Debras2019}. We will thus go beyond the classical models of tides that invoke inertial waves in the deep convective envelope \citep{Ogilvie2004}, viscoelastic tides in the rocky/icy core \citep{Remus2012,Remus2015}, or the combination of both \citep{Guenel2014}, and instead move towards models that consider gravito-inertial waves propagating in giant planets' interiors where both convective and stably stratified layers co-exist \citep{Andre2017, Andre2019, Pontin2020, Pontin2023, Lin2023, Dewberry2023}.

In this study, we develop a method to calculate the dissipation of the dynamical tidal response of a  self-gravitating, rotating fluid body composed of alternating convective layers and stably stratified layers and which takes into account the viscous, thermal and chemical dissipation processes. This is the first time that global models incorporate the consideration of all three dissipation mechanisms, as opposed to solely focusing on viscosity in previous models. The Coriolis force is fully taken into account, but the centrifugal force is neglected as a first step. This method allow us to compute the imaginary part of the tidal Love number for a given planetary interior model. We will focus specially on the latest Jupiter interior model constrained by Juno data and calculated by \cite{Debras2019}.

The paper is structured as follows: In Section \ref{sec:2}, we derive the model that allows us to study the dissipation of tidally forced waves. We provide a detailed explanation for the separation of equilibrium tides and dynamical tides. Additionally, we derive the energy equation of tidal flows within this framework.
In Section \ref{sec:3}, we present the Jupiter’s interior model used in this study.
Moving on to Section \ref{sec:4}, we outline our methodology for calculating the equilibrium tide in the adiabatic case. Then, we focus on computing the dynamical (wave-like) tide from 2D linear pseudo-spectral numerical simulations, which allow us to derive the associated dissipation. We then present, in Section \ref{sec:5}, the novel results obtained of tidal waves in Jupiter. Specifically, we cover the simultaneous inclusion of inertial waves in convection zones and gravito-inertial waves in stably stratified zones along with the evaluation of the dissipation resulting from the different dissipative processes. Finally, we summarise, in Section \ref{sec:6}, the key findings and implications of our study.

\section{Modelling tidally forced waves in giant planet interiors}\label{sec:2}
We study the linear excitation of (gravito-)inertial waves by an external tidal body forcing $\vec{F}_{\rm forcing}$. These waves are subject to dissipative processes, namely viscosity, thermal diffusion, and molecular diffusion (we assume that these diffusivities are uniform (cf. \ref{subsec:tranp_prop})).

\subsection{Governing equations}
We begin by writing the system of dynamical equations formed by the following set of equations. First, we write the continuity equation :
\begin{equation}\label{eq:continuity_g}
    D_t \rho+\rho \vec{\nabla} \cdot \vec{V}=0,
\end{equation}
where $\rho$ is  the density, $\vec{V}$ is the velocity field,  and $D_{t}=\partial_{t}+(\vec{V} \cdot \vec{\nabla})$ is the Lagrangian derivative. Then, we introduce the momentum equation :
\begin{equation}\label{eq:momentum_g}
    \rho \frac{D \vec{V}}{D t}=-\nabla P - \rho\nabla \Phi +\rho \nu \left(\vec{\nabla}^2 \vec{V}+\frac{1}{3} \nabla \nabla \cdot \vec{V}\right) + \vec{F}_{\rm forcing},
\end{equation}
where $P$ is the pressure, $\Phi$ the gravitational potential, $\nu$ is the kinematic viscosity assumed to be constant, and $\vec{F}_{\rm forcing}$ the tidal forcing. We adopt here the Stokes hypothesis, where the bulk viscosity is neglected.
We introduce also the heat (energy) equation :
\begin{equation}\label{eq:energy_g}
    \rho T D_t s= k \nabla^2 T,
\end{equation}
where $T$ is the temperature, $k$ is the thermal conductivity, and $s$ is the specific entropy such as $\displaystyle{\mathrm{d}s \coloneqq c_p\left(\frac{\mathrm{d}T}{T} - \nabla_{\mathrm{ad}} \frac{\mathrm{d}P}{P}\right)}$ with $\nabla_{\mathrm{ad}}\coloneqq\left(\frac{\mathrm{d} \ln T}{\mathrm{~d} \ln P}\right)_{s}$ the adiabatic temperature gradient and $c_p$ the specific heat capacity. We neglect here the viscous heating term and suppose $k$ is constant.
The chemical composition equation can be written as :
\begin{equation}\label{eq:chemical_g}
    D_t \mu= D_\mu \nabla^2 \mu,
\end{equation}
where $\mu$ is the molecular weight and $D_\mu$ is the molecular diffusion supposed constant.
The Poisson equation reads :
\begin{equation}\label{eq:Poisson_g}
    \nabla^2 \Phi=4 \pi G \rho,
\end{equation}
where $G$ the universal gravitational constant.
Finally, the general differential form of the equation of state \citep{Kippenhahn1994} is defined by :
\begin{equation}\label{eq:state_g}
    \frac{\partial \rho}{\rho} =\alpha \frac{\partial P}{P} - \delta \frac{\partial T}{T} + \phi \frac{\partial \mu}{\mu},
\end{equation}
with
\begin{equation}
     \alpha \coloneqq \left(\frac{\partial \ln \rho}{\partial \ln P}\right)_{T,\,\mu}, \; \delta\coloneqq-\left(\frac{\partial \ln \rho}{\partial \ln T}\right)_{P,\,\mu}, \; \phi\coloneqq\left(\frac{\partial \ln \rho}{\partial \ln \mu}\right)_{P,\,T}.
\end{equation}

\subsection{Linearisation}
We linearise the hydrodynamic system (Eqs.\,\ref{eq:continuity_g}-\ref{eq:state_g}) around the hydrostatic steady-state. Each scalar field $X\coloneqq\{P, \rho, \Phi, T, \mu\}$ is expanded as the sum of its hydrostatic value $X_0$ and of the Eulerian perturbations associated with the tides $X^\prime$:
\begin{equation}
    X(r,\theta,\varphi, t)=X_0(r)+{X}^\prime(r,\theta,\varphi, t).
\end{equation}
We neglect here the non-spherical character of the hydrostatic background due to the deformation associated with the centrifugal acceleration and the associated perturbation of the gravitational potential since gravito-inertial waves are only slightly affected by the deformation \citep[e.g.][]{Ballot2010, Dhouib2021a}. This implies that the background is independent of $\theta$, so $X_0=X_0(r)$.
We can write the   velocity  field, $\vec{V}$, as  the sum of the large-scale azimuthal velocity associated with the uniform rotation \citep[as a first step we neglect the differential rotation, since Jupiter's relative differential rotation is 4\%,][]{Guillot2018}, $\Omega$, and of the wave velocity, $\vec{v}$:
\begin{equation}
    \vec{V}(r,\theta,\varphi, t)= r \sin{\theta} \, \Omega \,  \vec{e}_{\varphi} + \vec{v}(r,\theta,\varphi, t),
\end{equation}
where $t$ is time and $(r,\theta,\varphi)$ are the usual spherical coordinates with their associated unit vector basis $(\vec{e}_r, \vec{e}_\theta, \vec{e}_\varphi)$. In this case, the linearised system (\eqref{eq:continuity_g}-\eqref{eq:state_g}) can be rewritten in the rotating frame as 
\begin{equation}
    \frac{\partial \rho^{\prime}}{\partial t} +\vec\nabla\cdot\left(\rho_{0} \vec{v}\right)=0,\label{eq:continuity_l}
\end{equation}
\begin{multline}
        \frac{\partial \vec{v}}{\partial t}+2 \vec{\Omega} \times \vec{v}= -\vec{\nabla} W- \frac{\vec{\nabla}\rho_0}{\rho_0}W^{\prime} +\frac{\rho^{\prime}}{\rho_{0}} \vec{g}_0 - \nabla \Phi^\prime \\ 
        + \nu \left(\vec{\nabla}^2 \vec{v}+\frac{1}{3} \nabla \nabla \cdot \vec{v}\right)  + \vec{F}_{\rm forcing},
\end{multline}
\begin{gather}
   \frac{\partial T^{\prime}}{\partial t}+\vec{v} \cdot \vec{\nabla} T_0 - \nabla_{\rm ad} \frac{T_0}{P_0} \left(\frac{\partial P^{\prime}}{\partial t}+\vec{v} \cdot \vec{\nabla} P_0 \right)= \kappa \vec{\nabla}^2 T^{\prime}, \label{eq:energy_l}\\
    \frac{\partial \mu^{\prime}}{\partial t}+\vec{v} \cdot \vec{\nabla} \mu_{0}=D_\mu \vec{\nabla}^2 \mu^{\prime},\label{eq:chemical_l}\\
    \nabla^2 \Phi^{\prime} = 4 \pi G \rho^{\prime},\\
    \frac{\rho^{\prime}}{\rho_0} =\alpha \frac{ P^{\prime}}{P_0} - \delta \frac{T^\prime}{T_0} + \phi \frac{ \mu^{\prime}}{\mu_0},
\end{gather}
where $ W^{\prime}=P^{\prime}/\rho_{0}$ is the normalised pressure, $\kappa=k/\rho_0 c_p$ is the thermal diffusivity supposed constant, and $\vec{g}_0=-\vec{\nabla}\Phi_0 = \vec{\nabla}P_0/\rho_0 = - g_0\vec{e}_r$ is the gravitational acceleration.

\subsection{Non-wavelike and wavelike tides}
We decompose the fluctuations associated with the tides into non-wavelike and wavelike parts
\begin{equation}
    Y = Y^{\rm nw} + Y^{\rm w},
\end{equation}
with $Y\coloneqq\{v_r, v_\theta, v_\varphi, X^{\prime}\}$ where $ Y^{\rm nw}$ is the non-wavelike (equilibrium) tide that satisfies the hydrostatic equilibrium \citep{Zahn1966, Zahn1989} and $Y^{\rm w}$ the wavelike (dynamical) tide that describes the propagation of waves \citep{Zahn1975, Ogilvie2004}.

\subsubsection{Non-wavelike part}
We assume that the non-wavelike part is adiabatic ($\alpha \approx 1/\Gamma_1$ and $\kappa =0$), where $\Gamma_{1}=(\partial \ln P_0 / \partial \ln \rho_0)_{s}$ is the first adiabatic exponent, and non-dissipative ($\nu=D_\mu=0$).
The planet is assumed to be tidally forced by a single potential component
$\vec{F}_{\rm forcing} = -\nabla \Psi$ where
\begin{equation}
    \Psi(r,\theta,\varphi, t)=\Psi_\ell(r) Y_\ell^m(\theta, \varphi) \mathrm{e}^{-\mathrm{i} \omega_0 t},\; \text{with } \Psi_\ell(r) = A \left(\frac{r}{R}\right)^\ell,
\end{equation}
where
\begin{equation}\label{eq:exp_A}
    A \propto \frac{G R^\ell M_{\rm satellite}}{a^{\ell+1}}
\end{equation}
is the forcing amplitude where $M_{\rm satellite}$ is the mass of the satellite and $a$ is the semi-major axis. In our linear numerical calculations, we use a normalised value of $A$ so we will set $A=1$. $Y_\ell^m(\theta, \varphi)$ is an orthonormalised spherical harmonic of degree $\ell$ and order $m$ and $\omega_0 = n\Omega_{\rm orbital}$ the tidal frequency in the inertial frame centred on the planet ($n$ labels temporal harmonics of the orbital motion and $\Omega_{\rm orbital}$ denotes the orbital frequency). We will only consider the dominant quadrupolar tidal component $\ell = m = 2$.

If we suppose that the adiabatic equilibrium tide is stationary in the frame rotating with the fluid inside the planet \citep{Remus2012}, we simplify the linearised heat (Eq.\,\ref{eq:energy_l}) and chemical composition (Eq.\,\ref{eq:chemical_l}) equations by neglecting $\partial P^{\rm nw}/\partial t$. In that case, the system of equations that describes the non-wavelike tides can be written as
\begin{gather}
    \rho^{\rm nw}+\vec\nabla\cdot\left(\rho_{0} \vec{\xi}^{\rm nw} \right)=0, \label{eq:continuity_nw}\\
    -\vec{\nabla} W^{\rm nw}  - \frac{\vec{\nabla}\rho_0}{\rho_0}W^{\rm nw}  +\frac{\rho^{\rm nw} }{\rho_{0}} \vec{g}_0- \nabla \Phi^{\rm nw} -\nabla \Psi = 0, \label{eq:momentum_nw}\\
     T^{\rm nw} +\xi_{r}^{\rm nw} T_0 \frac{N_{\rm t}^2}{g_0\delta}=0, \label{eq:energy_nw}\\
    \mu^{\rm nw}  - \xi_{r}^{\rm nw} \mu_0 \frac{N_\mu^2}{g_0\phi} =0,  \label{eq:chemical_nw} \\
    \nabla^2 \Phi^{\rm nw} = 4 \pi G \rho^{\rm nw}, \label{eq:poisson_nw}\\
    \frac{\rho^{\rm nw} }{\rho_0} = \frac{1}{\Gamma_1}\frac{P^{\rm nw}}{P_0}- \delta \frac{T^{\rm nw} }{T_0}+\phi \frac{\mu^{\rm nw} }{\mu_0}\label{eq:state_nw},
\end{gather}
where $\vec{\xi}^{\rm nw} $ is the displacement defined as $ \vec{v}^{\rm nw}= \partial\vec{\xi}^{\rm nw}/\partial t $ (it is customary to consider the displacement $\vec{\xi}^{\rm nw} $ instead of the velocity $\vec{v}^{\rm nw} $ in the calculation of the non-wavelike tide since it is a deformation induced by mass redistribution), 
\begin{equation}
    N_{\rm t}^2= - g_0\delta \frac{\mathrm{d} \ln P_0}{\mathrm{d} r} \left(\nabla_{\mathrm{ad}}-\frac{\mathrm{d} \ln T_0}{\mathrm{d} \ln P_0}\right),
\end{equation}
is the thermal Brunt-Väisälä frequency squared and
\begin{equation}
    N_\mu^2= -g_0 \phi \frac{\mathrm{d} \ln \mu_0}{\mathrm{d} r},
\end{equation}
is the compositional Brunt-Väisälä frequency squared.
The sum of these two qualities gives us the total Brunt–Väisälä frequency squared:
 \begin{equation}
    N^2=N_{\rm t}^2 + N_{\mu}^2=-g_0\left(\frac{1}{\rho_0} \frac{\mathrm{d} \rho_0}{\mathrm{d} r}-\frac{1}{\Gamma_{1} P_0} \frac{\mathrm{d} P_0}{\mathrm{d} r}\right).
    \end{equation}
If we write the Eq.\,\eqref{eq:momentum_nw} as:
\begin{equation}
    -\vec{\nabla} \left(W^{\rm nw} + \Phi^{\rm nw} +\Psi\right)  + \left(\rho^{\rm nw} - \frac{\mathrm{d}\rho_0}{\mathrm{d}P_0}P^{\rm nw} \right) \frac{\vec{\nabla} P_0}{\rho_0^2} = 0,
\end{equation}
we deduce that \citep{Ogilvie2014} 
\begin{gather}
    P^{\rm nw} = -\rho_0\left(\Phi^{\rm nw} + \Psi\right),\\
    \rho^{\rm nw} = \frac{\mathrm{d}\rho_0}{\mathrm{d}P_0}P^{\rm nw}.
\end{gather}
The non-wavelike gravitational potential is obtained by solving the Poisson's equation (Eq.\,\ref{eq:poisson_nw}) which can be rewritten as 
\begin{equation}\label{eq:poiss_solve}
    \frac{1}{r^2}\frac{\mathrm{d}}{\mathrm{d} r}\left(r^2\frac{\mathrm{d}\Phi^{\rm nw}_\ell}{\mathrm{d} r}\right) - \frac{\ell(\ell+1)}{r^2}\Phi^{\rm nw}_\ell + 4\pi G \frac{\mathrm{d}\rho_0}{\mathrm{d}P_0}\rho_0 \left(\Phi^{\rm nw}_\ell + \Psi_\ell\right) = 0,
\end{equation}
where 
\begin{equation}
    \Phi^{\rm nw}(r,\theta,\varphi, t)=\Phi^{\rm nw}_\ell(r) Y_\ell^m(\theta, \varphi) \mathrm{e}^{-\mathrm{i} \omega_0 t},
\end{equation}
and with the following boundary conditions to ensure its regularity near the centre ($r=\eta$, where $\eta$ is the aspect ratio) and its continuity at the surface ($r=1$) \citep[e.g.][]{Ogilvie2009}:
\begin{gather}
\frac{\mathrm{d} \ln \Phi^{\mathrm{nw}}_\ell}{\mathrm{d} \ln r}=\ell \;\text {at}\; r= \eta R, \label{eq:BD_phi_0}\\
\frac{\mathrm{d} \ln \Phi^{\mathrm{nw}}_\ell}{\mathrm{d} \ln r}=-(\ell+1) \; \text {at}\; r = R\label{eq:BD_phi_1}.
\end{gather}
By using Eqs.\,\eqref{eq:energy_nw}\,\&\,\eqref{eq:chemical_nw} in Eq.\,\eqref{eq:state_nw} we obtain
\begin{equation}\label{eq:energy_approx}
    \frac{1}{\Gamma_1}\frac{P^{\rm nw}}{P_0} - \frac{\rho^{\rm nw} }{\rho_0} + \frac{N^2}{g_0} \xi_r^{\rm nw} = 0,
\end{equation}
where $\vec{\xi}^{\rm nw} = \xi_r^{\rm nw}  \vec{e}_r+ \vec{\xi}^{\rm nw}_h$ such that $\vec{\xi}_h^{\rm nw}\cdot\vec{e}_r=0$. From this equation we can derive the expression for the non-wavelike radial displacement 
\begin{equation}
    N^2\left(\xi_{r, \ell}^{\rm nw}+\frac{\Phi^{\rm nw}_\ell + \Psi_\ell}{g_0}\right)=0,
\end{equation}
so when $N^2\neq 0$ we obtain
\begin{equation}\label{eq:xinw_r}
    \xi_{r, \ell}^{\rm nw} = - \frac{\Phi^{\rm nw}_\ell + \Psi_\ell}{g_0}.
\end{equation}
Subsequently, from Eq.\,\eqref{eq:continuity_nw} we can derive the non-wavelike horizontal displacement
\begin{equation}\label{eq:xinw_h}
    \xi_{h, \ell}^{\rm nw} = \frac{1}{\ell(\ell+1)}\left(2\xi_{r, \ell}^{\rm nw} + r \frac{\mathrm{d} \xi_{r, \ell}^{\rm nw}}{\mathrm{d} r}\right),
\end{equation}
therefore we can deduce that 
\begin{equation}
    \operatorname{div}\vec{\xi}^{\rm nw} = 0.
\end{equation}
This is the conventional equilibrium tide \citep{Zahn1966,Zahn1989,Remus2012}.
This solution applies not only to stably stratified zones but also to convective regions since $N^2$ is not equal to zero but slightly negative and the fact that we generally set $N^2$ in these zones to zero is only an approximation. Thus, Eqs.\,\eqref{eq:xinw_r}\,\&\,\eqref{eq:xinw_h} may be applied to the whole fluid domain inside the planet.
\citet{Terquem1998} and \citet{Goodman1998} argued that this equilibrium tide solution does not apply to convective regions, as they assumed that the convective zone is adiabatically stratified ($N^2=0$). A comparison between these two definitions of the non-wavelike tides performed by \citet{Barker2020} highlights the fact that in the interface between convective zones and stably stratified zones, a discontinuity arises in the horizontal component of displacement. This situation poses a problem both physically, since the ellipsoidal deformation and the related displacement has no reason to be discontinuous at convective-radiative boundaries, and numerically when dealing with multi-zone problems. We can therefore use the solution derived by \citet{Zahn1966,Zahn1989} and \citet{Remus2012} which applies in stably stratified zones, but also in convective regions, since $N^2$ is not strictly equal to zero, but slightly negative.

\subsubsection{Wavelike part}
To derive the wavelike part, we assume first the Boussinesq approximation \citep{Spiegel1960} which neglects the density variations except where they appear in the buoyancy term, so the acoustic waves are filtered out. 
This approximation is an essential first step for addressing such a complex problem where the eigenmodes at these low frequencies are generally singular and are regularised by diffusion processes (see Sec.\,\ref{sec:6} for the discussion on the use of the anelastic approximation instead of the Boussinesq one).
In fact, calculating inertial and gravito-inertial waves in an internal structure model with multiple transition layers poses a challenge, particularly given the presence of the strong density gradients. Additionally, incorporating three diffusion processes with coefficients spanning several orders of magnitude, potentially reaching very low values, adds another layer of complexity that is numerically demanding. To manage these complexities effectively, it is necessary to start with a simplified model to control the physical processes before moving on to the following stages. This approach is crucial for acknowledging and addressing possible biases introduced during the analysis. In Sec.\,\ref{sec:6}, we will discuss carefully the potential limitations of this first necessary study within the Boussinesq approximation and the needs to go beyond it in a near future.
Then, we assume the Cowling approximation \citep{Cowling1941} which neglects the perturbations of the gravitational potential induced by the waves since the perturbations induced by the non-wavelike tides are dominant \citep[e.g.][]{Ogilvie2004}. The system of equations that describes the wavelike tides can thus be written as
\begin{gather}
    \vec\nabla\cdot \vec{v}^{\rm w} =0, \\
    \frac{\partial \vec{v}^{\rm w} }{\partial t}+2 \vec{\Omega} \times \vec{v}^{\rm w} =-\vec{\nabla}W^{\rm w}   + g_{0} \left(\delta  \frac{T^{\rm w} }{T_0} - \phi  \frac{\mu^{\rm w} }{\mu_0} \right)\vec{e}_r + \nu \vec{\nabla}^2 \vec{v}^{\rm w} + \vec{f}^{\rm nw}, \\
    \frac{\partial T^{\rm w} }{\partial t}+v_{r}^{\rm w} T_0 \frac{N_{\rm t}^2}{g_0\delta}=\kappa \vec{\nabla}^2  T^{\rm w}, \\
    \frac{\partial \mu^{\rm w} }{\partial t} - v_{r}^{\rm w} \mu_0 \frac{N_\mu^2}{g_0\phi} =D_\mu \vec{\nabla}^2  \mu^{\rm w},  \\
    \frac{\rho^{\rm w} }{\rho_0} = - \delta \frac{T^{\rm w} }{T_0}+\phi \frac{\mu^{\rm w} }{\mu_0},
\end{gather}
with 
\begin{equation}\label{eq:foricng}
    \vec{f}^{\rm nw} = - \frac{\partial \vec{v}^{\rm nw} }{\partial t} - 2 \vec{\Omega} \times \vec{v}^{\rm nw},
\end{equation}
the forcing term which arises when solving the non-wavelike tides as a residual force, as the non-wavelike does not satisfy the equation of motion due to the omission of inertial forces associated with this flow. This force encompasses the acceleration of the non-wavelike tide and the Coriolis acceleration applied to it, and will force the gravito-inertial tidal waves \citep[see also][]{Ogilvie2005, Andre2019}.

\paragraph{Non-dimensional system:}
We choose the planet’s radius $R$ for the length scale and $(2\Omega)^{-1}$ for the timescale ($t=\tau/2\Omega$). Therefore, we can define the normalised quantities as follows $\vec{v}^{\rm w}=2 \Omega R \vec{u}$,
$W^{\rm w}=4 \Omega^{2} R^{2} \Pi$, $g_0=4\Omega^2R g_0^*$, $\vec{f}^{\rm nw}=4\Omega^2R\vec{f}^*$, and the normalised differential operator as $R\vec{\nabla}=\vec{\nabla^*}$. 
We write the normalised temperature and chemical composition as
$T^{\rm w}= T_{\rm i} \Theta$,  $\mu^{\rm w}=\mu_{\rm i}M$, $T_0=T_{\rm i}T_0^*$ and $\mu_0=\mu_{\rm i}\mu_0^*$, where $T_{\rm i}$ and $\mu_{\rm i}$ are the temperature and the molecular weight respectively at the inner boundary.
So, we can write the normalised system as
\begin{gather}
    \vec{\nabla^*}\cdot \vec{u} =0, \label{eq:continuty}\\
    \frac{\partial \vec{u}}{\partial \tau}+\vec{e}_{z} \times \vec{u}=-\vec{\nabla} \Pi + \left(\frac{\delta }{T_0^*}\Theta-\frac{\phi }{\mu_0^*}M\right) g_0^* \vec{e}_r +\mathrm{E} \vec{\nabla^*}^2 \vec{u} + \vec{f}^*,\label{eq:momentim_norm}\\
    \frac{\partial \Theta}{\partial \tau}+  \frac{{T_0^* N_{\rm t}^*}^{2}}{g_0^* \delta}u_{r}= \frac{\mathrm{E}}{\mathrm{Pr}} \vec{\nabla^*}^2 \Theta, \label{eq:energy_norm}\\
    \frac{\partial M}{\partial \tau} - \frac{{\mu_0^* N_\mu^*}^{2}}{g_0^* \phi} u_{r} =\frac{\mathrm{E}}{\mathrm{Sc}} \vec{\nabla^*}^2 M, \label{eq:chemical_norm}
\end{gather}
where we have defined the normalised Brunt–Väisälä frequencies
\begin{equation}
{N_{\rm t}^*}^{2}=\frac{N_{\rm t}^2}{4 \Omega^{2}} \text{  and  }
{N_\mu^*}^{2}=\frac{N_\mu^2}{4 \Omega^{2}}. 
\end{equation}
These equations are governed by three dimensionless numbers :
the Prandtl number defined as the ratio of the kinematic viscosity ($\nu$) to the thermal diffusivity ($\kappa$) :
\begin{equation}
    \mathrm{Pr}=\frac{\nu}{\kappa},
\end{equation}
the Schmidt number defined as the ratio of the kinematic viscosity ($\nu$) to the molecular diffusivity ($D_\mu$) :
\begin{equation}
    \mathrm{Sc}=\frac{\nu}{D_\mu},
\end{equation}
and the Ekman number which compares the ratio between the viscous force and the Coriolis force :
\begin{equation}
    \mathrm{E}=\frac{\nu}{2 \Omega R^{2}}.
\end{equation}
\subsection{Energy equation}

From the scalar product between $\Bar{\vec{u}}$ (where $\Bar{\square}$ denotes the complex conjugate) and the momentum equation \eqref{eq:momentim_norm} and by using Eqs.\,\eqref{eq:energy_norm}\,\&\,\eqref{eq:chemical_norm}, we obtain the energy equation 
\begin{equation}
    \partial_\tau \left(E_{\mathrm{k}} + E_{\mathrm{p}, \mathrm{th}} + E_{\mathrm{p},\mu}\right) = P_{\mathrm{acou}} + D_{\mathrm{th}} + D_{\mathrm{ch}} + D_{\mathrm{visc}} + P_{\rm tide},
\end{equation}
with 
\begin{equation}
    E_{\mathrm{k}}= \frac{1}{2} ||\vec{u}||^2
\end{equation}
the specific kinetic energy,
\begin{equation}
    E_{\mathrm{p}, \mathrm{th}}  = \left\{
    \begin{array}{ll}
        \displaystyle{\frac{1}{2}\left(\frac{\delta g_0^*}{T_0^*N_{\rm t}^*}\right)^2 {|\Theta|}^2} & \mbox{if } N_{\rm t}^{2} \neq 0  \\
        0 & \mbox{if not}
    \end{array}
\right.
\end{equation}
the specific potential energy associated with thermal stratification,
\begin{equation}
    E_{\mathrm{p}, \mu}  = \left\{
    \begin{array}{ll}
        \displaystyle{\frac{1}{2}\left(\frac{\phi g_0^*}{\mu_0^* N_\mu^*}\right)^2 {|M|}^2 } & \mbox{if } N_\mu^{2} \neq 0  \\
        0 & \mbox{if not}
    \end{array}
\right.
\end{equation}
the specific potential energy associated with chemical stratification,
\begin{equation}
    P_{\mathrm{acou}} = - \vec{\nabla^*}\Pi \cdot \Bar{\vec{u}}
\end{equation}
the specific work of pressure forces which can be related to the acoustic flux $\Vec{\nabla}\cdot\left(P^\prime \vec{u}\right)$,
\begin{equation}
    D_{\mathrm{th}}  = \left\{
    \begin{array}{ll}
        \displaystyle{\frac{\mathrm{E}}{\mathrm{Pr}} \left(\frac{\delta g_0^*}{T_0^* N_{\rm t}^*}\right)^2  \Bar{\Theta} \vec{\nabla^*}^2 \Theta }  & \mbox{if } N_{\rm t}^{2} \neq 0  \\
        0 & \mbox{if not}
    \end{array}
\right.
\end{equation}
the specific power dissipated by thermal diffusion, 
\begin{equation}
    D_{\mathrm{ch}}  = \left\{
    \begin{array}{ll}
        \displaystyle{\frac{\mathrm{E}}{\mathrm{Sc}}  \left(\frac{\phi g_0^*}{\mu_0^* N_\mu^*}\right)^2 \Bar{M} \vec{\nabla^*}^2 M } & \mbox{if } N_\mu^{2} \neq 0  \\
        0 & \mbox{if not}
    \end{array}
\right.
\end{equation}
the specific power dissipated by chemical diffusion,
\begin{equation}
   D_{\mathrm{visc}} =  E  \vec{\nabla^*}^2 \vec{u}   \cdot \Bar{\vec{u}}
\end{equation}
the specific power dissipated by viscous friction,
\begin{equation}
    P_{\rm tide} =\vec{f}^* \cdot \Bar{\vec{u}}
\end{equation}
the specific tidal power.
Then after spatial integration over the volume $V$, we obtain 
\begin{equation}\label{eq:energy_integ_V}
    \partial_t \left(\widetilde{E}_{\mathrm{k}} + \widetilde{E}_{\mathrm{p}, \mathrm{th}} + \widetilde{E}_{\mathrm{p},\mu}\right) = \widetilde{P}_{\mathrm{acou}} + \widetilde{D}_{\mathrm{th}} + \widetilde{D}_{\mathrm{ch}} + \widetilde{D}_{\mathrm{visc}} + \widetilde{P}_{\rm tide},
\end{equation}
where $\displaystyle{\widetilde{\square} = \frac{1}{2} \Re\left[\int_{V}  \square\mathrm{d}V \right]}$. Here, we assume that the density is constant by adopting the Boussinesq approximation, allowing us to simplify it in this equation on both sides.

\section{Interior model of Jupiter as revealed by Juno} \label{sec:3}
\subsection{Five-layer model}\label{subsec:struc_model}
\begin{figure}
    \centering
    \resizebox{\hsize}{!}{\includegraphics{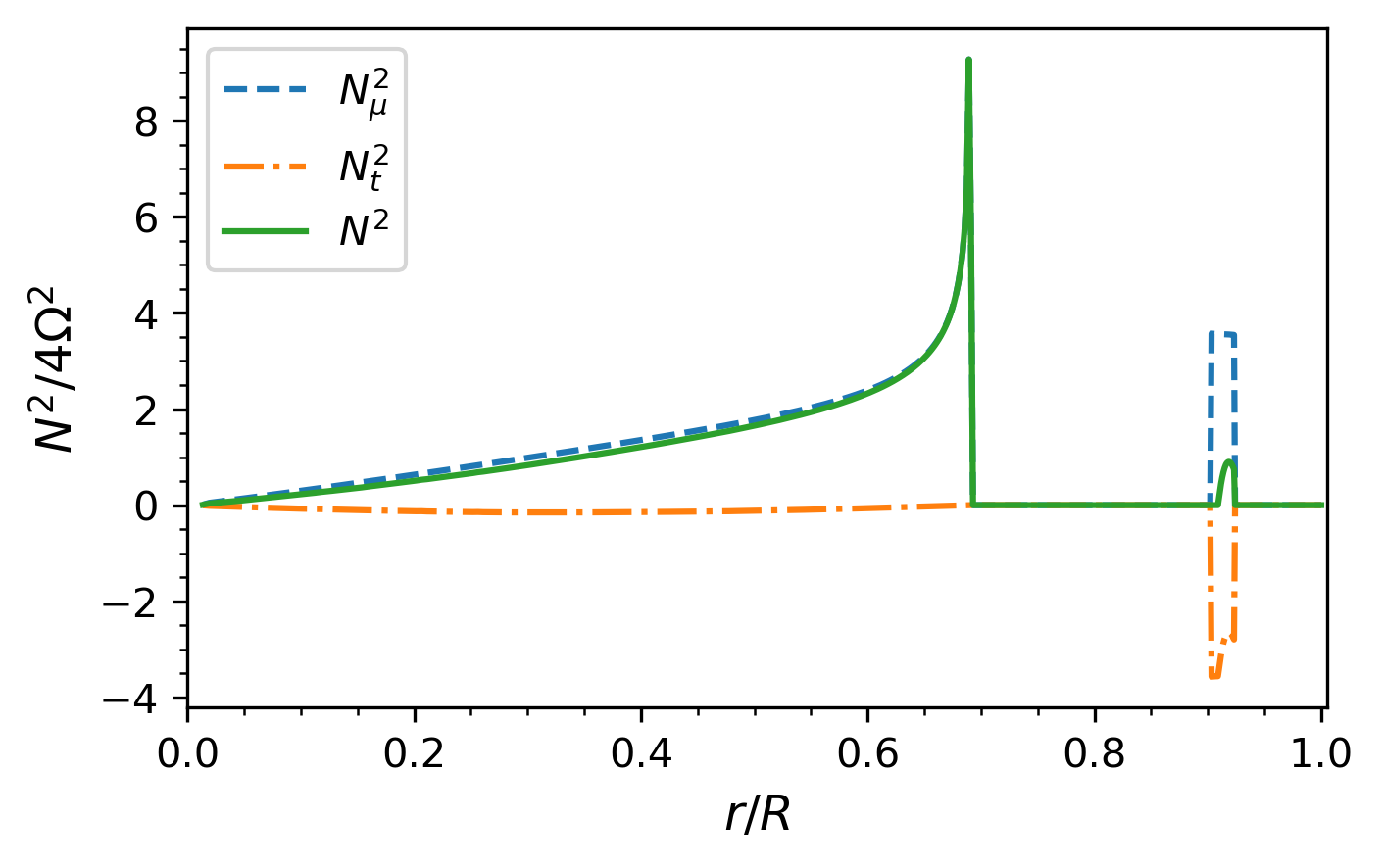}}
    \caption{Profiles of the compositional ($N_\mu^2$), thermal ($N_{\rm t}^2$), and total ($N^2$) Brunt–Väisälä frequencies squared normalized by the inertial frequencies squared ($4\Omega^2$) as a function of the normalised radius $r/R$.}
    \label{fig:freq_N}
\end{figure}

\begin{figure}
    \centering
    \resizebox{\hsize}{!}{\includegraphics{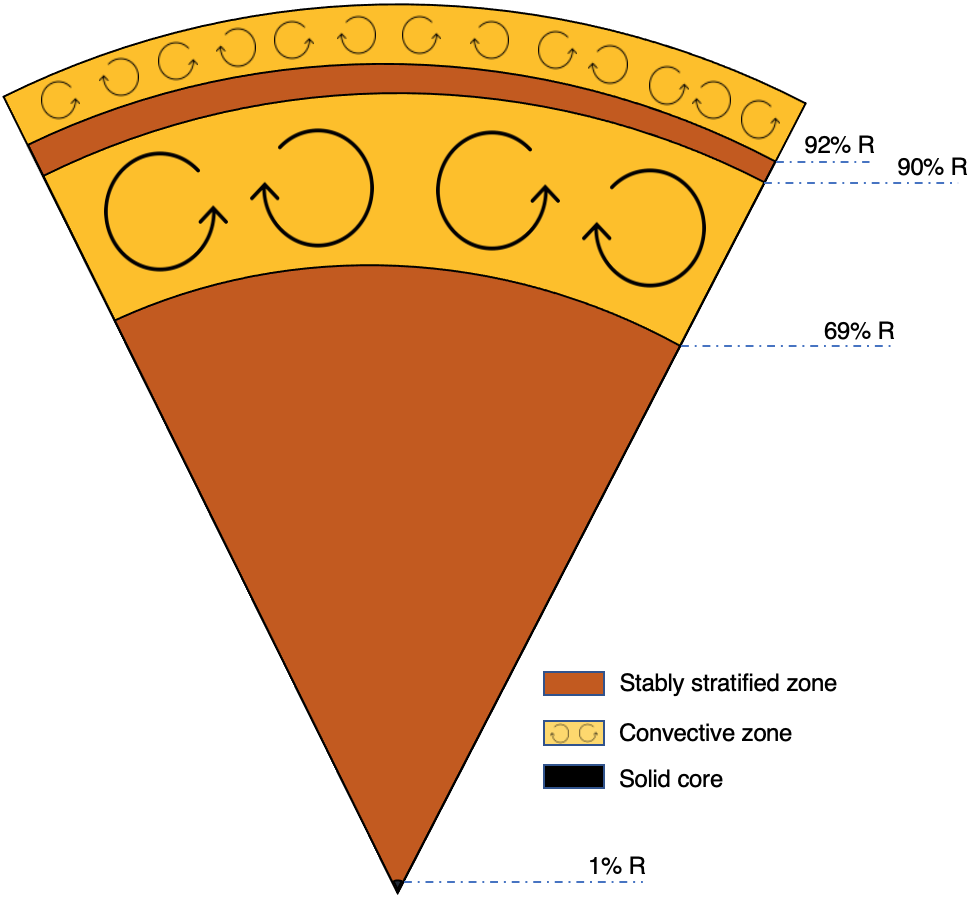}}
    \caption{Schematic of the model of Jupiter’s interior used in this study.}
    \label{fig:jup_model}
\end{figure}

In order to assess the dissipation of dynamical tides in Jupiter, we need to prescribe the background state profiles of the different quantities and the Brunt–Väisälä frequencies. 
Recent observations have made significant advancements in understanding the internal structure of gas giant plants, such as Jupiter and Saturn, yet some degrees
of uncertainties remains.
As we can see in Fig.\,\ref{fig:freq_N}, the internal structure model computed by \cite{Debras2019} to reproduce Jupiter's multipolar moments as measured by Juno, assumes an extended diluted core of radius $0.69R$ treated as a stably stratified fluid layer \citep[a similar zone is probably also present in Saturn;][]{Mankovich2021} and a convective envelope that features a small stably stratified layer between $0.9R$ and $0.92R$, possibly resulting from H-He immiscibility \citep{Debras2019}. The equation of state used to compute this model is the one derived in \cite{Chabrier2021}.
As illustrated in Fig.\,\ref{fig:jup_model}, starting from its surface and moving towards the core, Jupiter is thought to exhibit the following layers:
\begin{itemize}
    \item Gaseous envelope: this outermost layer is characterized by convective motion and differential rotation.
    \item Transitional stably stratified  zone: this region is considered to be potentially semi-convective. It is also known as a double-diffusive zone, as proposed by \cite{Leconte2012}.
    \item Internal convective zone: this layer is magnetised and composed of metallic hydrogen and helium, and it is rotating quasi-uniformly since if the rotation were significantly different from solid rotation, the ohmic dissipation would become inexplicable \citep[e.g.][]{Liu2008, Guillot2018}.
    \item Stably stratified zone : 
    this layer is located closer to the core and may exhibit double diffusion convection or a diluted core structure due to stabilizing composition gradients \citep{Leconte2012, Leconte2013, Wilson2012a, Wilson2012b,Wahl2013, Gonzalez2014,Mazevet2015, Wahl2017}.
    \item Potential unstable solid core of size $1.4\%$ of radius: made up of rock or ice. While this size may slightly vary based on different models, we are unable to construct a model with a substantial core, bigger than $\sim10\%$, that respect the Juno's constraints \citep[e.g.][]{Debras2019}.
\end{itemize}
Note that as a first step, we will not take into account in this study the differential rotation and magnetism.

\subsection{Transport properties}\label{subsec:tranp_prop}
\begin{figure}
\centering
\resizebox{\hsize}{!}{\includegraphics{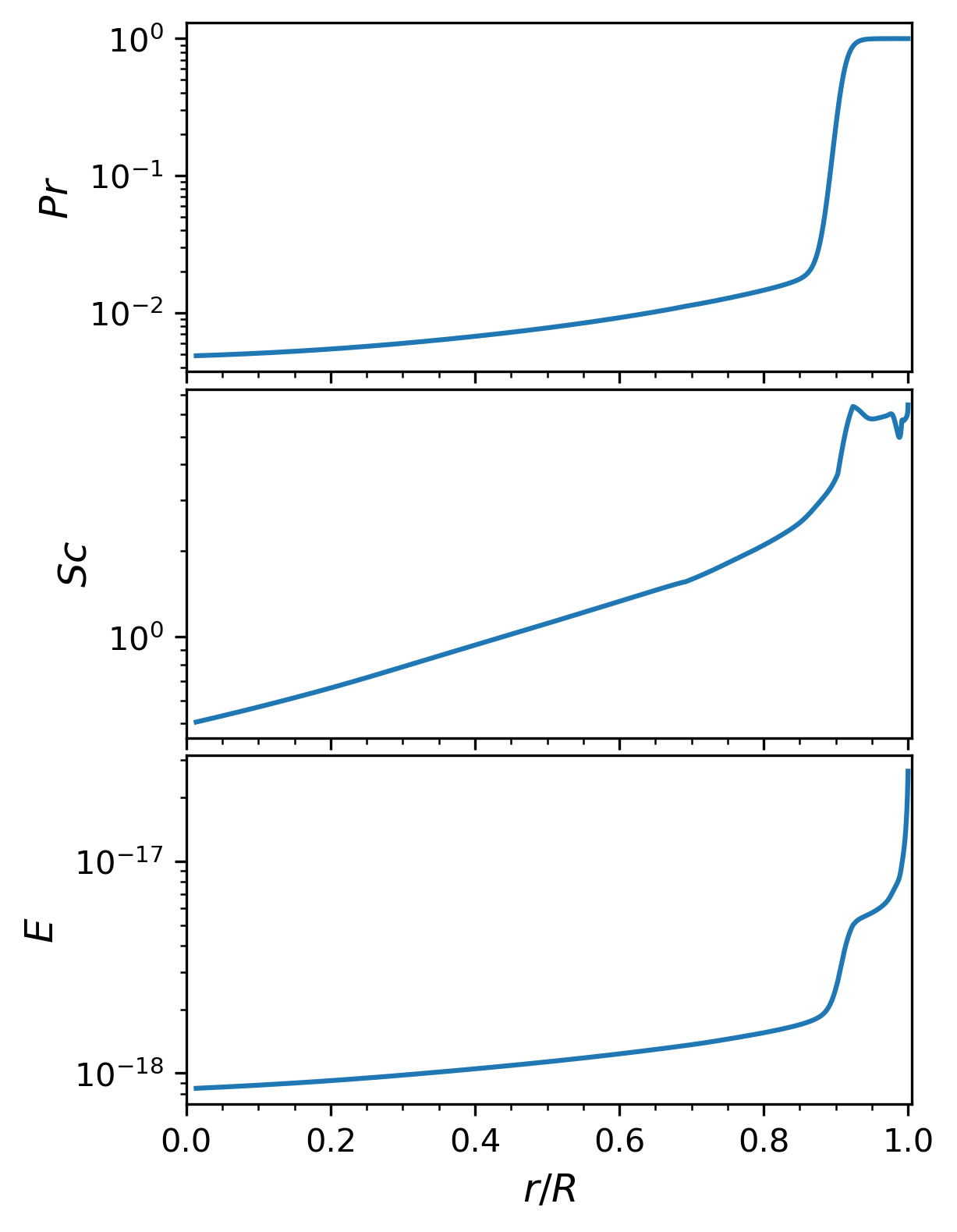}}
\caption{Prandtl, Schmidt, and Ekman number profiles as a function of the normalised radius based on \cite{Stevenson1977}.}
\label{fig:e_sc_pr}
\end{figure}

Using the transport properties outlined in \cite{Stevenson1977}, we perform calculations to determine the various molecular diffusivities within Jupiter and the associated dimensionless numbers (the expressions of these numbers are given in Appendix \ref{append:numb_exp}). The radial profile of these dimensionless numbers (Prandtl, Schmidt and Ekman) is displayed in Fig.\,\ref{fig:e_sc_pr}.
We can see that the influence of viscous forces is generally small compared to the Coriolis acceleration. Consequently, the Ekman number, which characterizes the ratio of viscous to Coriolis forces, becomes extremely low, typically on the order of $10^{-17}$ when assuming molecular viscosity. However, due to the challenges in accurately resolving shear layers at very low diffusivity levels, such Ekman number regimes cannot be reached numerically. The numerical strategy adopted is therefore to reach the lowest possible Ekman number values, hoping to have reached a regime where the scaling laws obtained will apply to the lower astrophysical values. Moreover, this Ekman number value is calculated with a  molecular viscosity value, whereas inertial tidal waves could in reality be subject to a turbulent effective viscosity \citep{Ogilvie2004, Ogilvie2007, Mathis2016, Duguid2020, Vidal2020, Vries2023} whose larger values would lead to a larger Ekman number.
Indeed, by employing the non-rotating mixing-length theory, we can make a rough estimation of the turbulent effective eddy viscosity in convective regions resulting in an Ekman number of approximately $10^{-7}$ \citep{Guillot2004}. When replacing the standard non-rotating mixing length theory by the rotating mixing length theory developed in \cite{Stevenson1979}, we end up \citep[following][]{Mathis2016} with a much smaller turbulent Ekman number of approximatively $10^{-15}$, closer to the microscopic value because of the inhibition of convection by rapid rotation \citep{Fuentes2023}.
On the other hand, the Prandtl number, which measures the ratio of viscosity to thermal diffusivity, is low within the planet, approximately $10^{-2}$, but increases to around 1 near the surface (at $r>0.9R$).
Regarding the Schmidt number, which characterizes the ratio of viscosity to molecular diffusivity, it remains close to unity throughout the planet. 
As a first step, we will assume in our simulations that these dimensionless numbers are constant. Then, we can study the impact of their variations by exploring the parameter domain.

\section{Numerical resolution}\label{sec:4}
Our attention in this paper will be directed towards the  $\ell= m = 2$ component of the tide, as it is commonly considered to be the most prominent for quasi-circular and coplanar two-body system \citep{Mathis&LePoncin2009, Ogilvie2014}.

\subsection{Solving the non-wavelike part}
\begin{figure*}
\centering
\resizebox{\hsize}{!}{\includegraphics{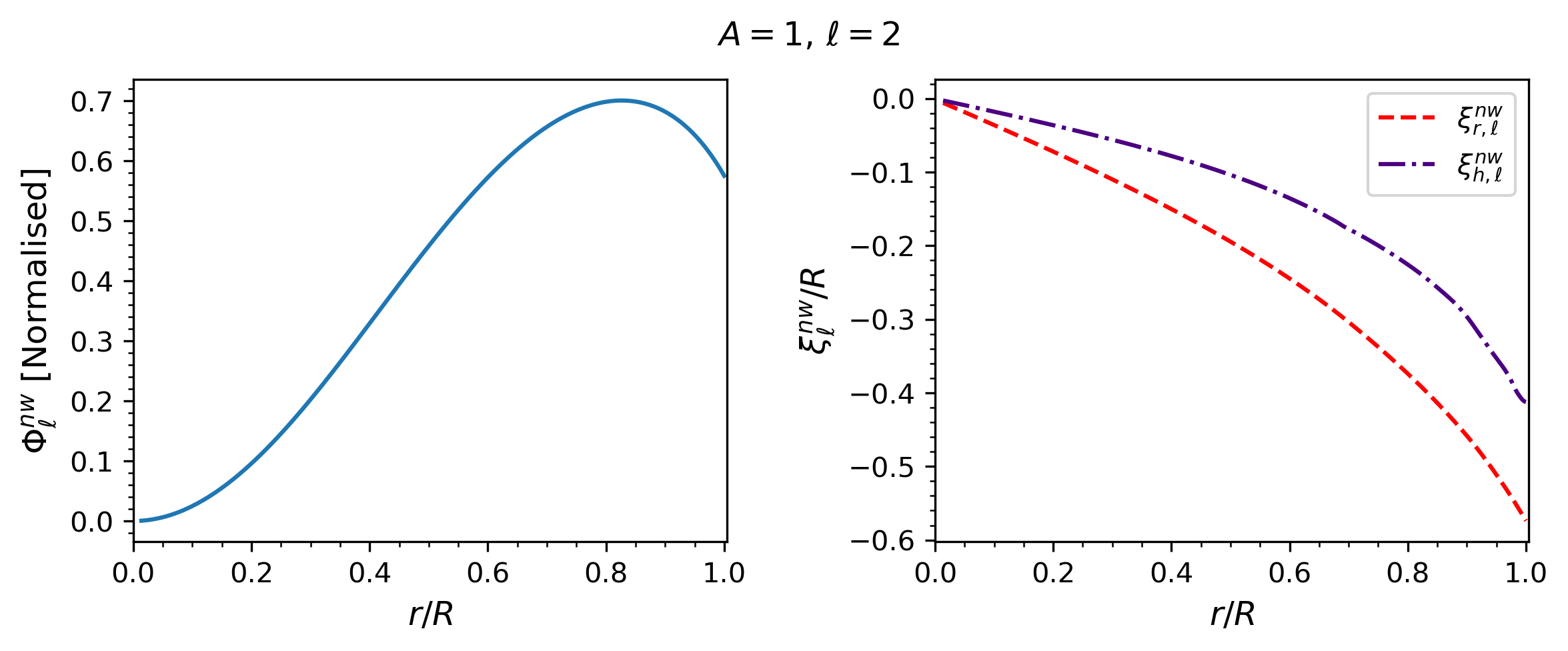}}
\caption{Non-wavelike (equilibrium) gravitational potential (left) and the corresponding  radial  and horizontal displacements (right) as a function of the normalised radius for $A=1$ and $\ell=2$.}
\label{fig:phi_xi_nw}
\end{figure*}

Our goal here is to find the non-wavelike displacement $\vec{\xi}^{\rm nw}$ as it is needed in the expression for the effective forcing driving dynamical tides (Eq.\,\ref{eq:foricng}). Therefore, we must first solve Poisson's equation (Eq.\,\ref{eq:poiss_solve}) numerically with boundary conditions \eqref{eq:BD_phi_0} and \eqref{eq:BD_phi_1} using the density and pressure background profiles ($\rho_0$ and $P_0$) that we compute using the structure model defined in Sec.\,(\ref{subsec:struc_model}). Then we can easily compute the vertical and horizontal non-wavelike displacement using Eqs.\,\eqref{eq:xinw_r}\,\&\,\eqref{eq:xinw_h}. We can see in Fig.\,\ref{fig:phi_xi_nw} these quantities as a function of the normalised radius for $A=1$ and $\ell=2$.
We emphasize here that setting the forcing value to $A=1$ gives very high displacement values, while a realistic forcing value (Eq.\,\ref{eq:exp_A}) would give much lower values. Since we are dealing here with the linear case, the choice of $A$ has no impact on our results (ultimately, we want to calculate the Love number defined just after in Eq.\,\eqref{eq:rek}, which is a ratio where $A$ will be simplified).

Once the solution to Poisson's equation (Eq.\,\ref{eq:poiss_solve}) is obtained numerically, the Love number \citep{Love1911} is readily given by
\begin{equation}\label{eq:rek}
    k_\ell^{\rm nw}=\frac{\Phi_\ell^{\rm nw}(r=R)}{\Psi_\ell(r=R)}.
\end{equation}
Note that this includes only non-wavelike tides, and it is real since we do not take into account its dissipation. For $\ell=2$ we find that $k_2^{\rm nw}=0.638$.
However, it is important to acknowledge that our assumption of a spherical planet does not hold true for Jupiter, as its rapid rotation causes it to be flattened, leading to a discrepancy between our value and the calculated values by \cite{Wahl2020}. In fact, they investigated the non-rotating case and found a value of $k_2^{\rm nw}=0.536$. However, when accounting for the planet's rotation, they determined that the value changes to $k_2^{\rm nw}=0.589$ for $m=2$. It is worth noting that the Love number becomes dependent on the azimuthal order $m$ when considering the effects of rotation.
Our approach takes into account rotation while neglecting flattening, whereas the method of \cite{Wahl2020} takes into account both rotation and induced deformation simultaneously. Since our main objective in this study is to understand wavelike tides, we compute the non-wavelike tides only to calculate the forcing term. Therefore, we can omit the flattening as a first step, considering that the forcing term will be only slightly modified.
The discrepancy of the calculated equilibrium Love number in comparison to the observed value \citep[$k_{22} = 0.565$,][]{Durante2020} suggests uncharacterised dynamical (wavelike) contribution due to tidal waves.

\subsection{Solving the wavelike part}
We expand the velocity, temperature, molecular weight and reduced pressure on spherical harmonics \citep{Rieutord1987} as
\begin{multline}
    \vec{u}=\sum_{\ell=0}^{+\infty} \sum_{m=-\ell}^{\ell} \left(u^{m}_{\ell}(r) \vec{R}_{\ell}^{m}(\theta, \varphi) + v^{m}_{\ell}(r) \vec{S}_{\ell}^{m}(\theta, \varphi)\right. \\
    \left.+ w^{m}_{\ell}(r) \vec{T}_{\ell}^{m} (\theta, \varphi) \right) \operatorname{e}^{-i\tilde{\omega} \tau}, \label{eq:u_expa}
\end{multline}
\begin{gather}
    \Theta=\sum_{\ell=0}^{+\infty} \sum_{m=-\ell}^{\ell} t^{m}_{\ell}(r) Y_{\ell}^{m}(\theta, \varphi)\operatorname{e}^{-i\tilde{\omega} \tau},\label{eq:t_expa}\\
    M=\sum_{\ell=0}^{+\infty} \sum_{m=-\ell}^{\ell} \mu^{m}_{\ell}(r) Y_{\ell}^{m}(\theta, \varphi)\operatorname{e}^{-i\tilde{\omega} \tau},\label{eq:mu_expa}\\
    \Pi = \sum_{\ell=0}^{+\infty} \sum_{m=-\ell}^{\ell} p^{m}_{\ell}(r) Y_{\ell}^{m}(\theta, \varphi)\operatorname{e}^{-i\tilde{\omega} \tau},\label{eq:p_expa} 
\end{gather}
with $\vec{R}_{\ell}^m=Y_{\ell}^m \vec{e}_r$,  $\vec{S}_{\ell}^m=\vec{\nabla} Y_{\ell}^m$, $\vec{T}_{\ell}^m=\vec{\nabla} \times \vec{R}_{\ell}^m$ and $\tilde{\omega}=\omega/2\Omega$ the normalised frequency associated with the tidal forcing in the rotating frame.
Then, the projection of the linearised dimensionless system (Sect.\,\ref{subsubsect:eq_to_solve}) and the associated  boundary conditions (Sect.\,\ref{subsubsect:bc}) are solved using the linear 2D pseudo-spectral code LSB \citep[Linear Solver Builder,][]{Valdettaro2007}.
These equations are discretised in the radial direction on the Gauss-Lobatto collocation nodes associated with Chebyshev polynomials. They are truncated at order $N_r$ for the Chebyshev basis and at order $N_\ell$ for the spherical harmonics basis. 
The governing equations \eqref{eq:solve1} to \eqref{eq:solvelast} and adopted boundary conditions (\ref{subsubsect:bc}) form a linear system of the form $\mathcal{M}X=\mathrm{F}$ which is solved on each point of the radial grid, given the value of the forcing frequency $\tilde{\omega}$ and azimuthal order $m$.

\subsubsection{System of equations to solve}\label{subsubsect:eq_to_solve}
Using equations (\ref{eq:u_expa}-\ref{eq:p_expa}) and the expressions of the operators in the
spherical harmonics basis specified in Appendix\,\ref{append:exp_harmo}, we can rewrite the system \eqref{eq:continuty}-\eqref{eq:chemical_norm} as
\begin{equation}
    \mathrm{d}_{r}u^{m}_{\ell} + \frac{2}{r}  u^{m}_{\ell} - \ell(\ell+1) \frac{v^{m}_{\ell}}{r} = 0,\label{eq:solve1}
\end{equation}
\begin{multline}
       \mathrm{E}\Delta_{\ell} u^{m}_{\ell}-\left(\frac{2\mathrm{E}}{r^{2}} - i \tilde{\omega}\right) u^{m}_{\ell} + \left(i m+ \frac{2\mathrm{E}}{r^{2}} \ell(\ell+1)\right) v^{m}_{\ell} - \\ \beta_{\ell-1}^{\ell} w_{\ell-1}^m - \beta_{\ell+1}^{\ell} w_{\ell+1}^m
        -\mathrm{d}_r p^{m}_{\ell}
         + \frac{\delta g_0^*}{T_0^*} t^{m}_{\ell}  - \frac{\phi  g_0^*}{\mu_0^*} \mu^{m}_{\ell}  = - f^{\ell,m}_{\rm R},
\end{multline}
\begin{multline}
       \mathrm{E} \Delta_{\ell} v^{m}_{\ell}+ \left(i\tilde{\omega} + \frac{i m}{\ell(\ell+1)} \right) v^{m}_{\ell} +\left(\frac{2\mathrm{E}}{r^{2}} + \frac{i m}{\ell(\ell+1)}\right) u^{m}_{\ell}
         - \\ \gamma_{\ell-1}^{\ell} w_{\ell-1}^m - \gamma_{\ell+1}^{\ell} w_{\ell+1}^m  -\frac{p^{m}_{\ell}}{r} =-f^{\ell,m}_{\rm S},
\end{multline}
\begin{multline}
       \mathrm{E} \Delta_{\ell} w^{m}_{\ell} +\left(i\tilde{\omega} + \frac{i m}{\ell(\ell+1)} \right)w^{m}_{\ell} + \gamma_{\ell-1}^{\ell} v_{\ell-1}^m + \gamma_{\ell+1}^{\ell} v_{\ell+1}^m - \\ 
       \frac{\alpha_{\ell-1}^{\ell}}{\ell} u_{\ell-1}^m + \frac{\alpha_{\ell+1}^{\ell}}{\ell+1} u_{\ell+1}^m = - f^{\ell,m}_{\rm T},
\end{multline}
\begin{equation}
     \frac{\mathrm{E}}{\mathrm{Pr}}\Delta_\ell t^{m}_{\ell}  - u^{m}_{\ell}  \frac{T_0^*{N_{\rm t}^*}^{2}}{g_0^* \delta} + i\tilde{\omega} t^{m}_{\ell}=0,
\end{equation}
\begin{equation}
     \frac{\mathrm{E}}{\mathrm{Sc}}\Delta_\ell \mu^{m}_{\ell}  + u^{m}_{\ell} \frac{\mu_0^*{N_\mu^*}^{2}}{g_0^* \phi} + i\tilde{\omega} \mu^{m}_{\ell}=0,\label{eq:solvelast}
\end{equation}
where
\begin{gather}
    \mathrm{d}_{r}\square=\frac{\mathrm{d}\square}{\mathrm{d} r},\\
    \mathrm{d}_{r^2}^{2}\square=\frac{\mathrm{d}^2\square}{\mathrm{d} r^2},\\
    \Delta_{\ell}\square=\mathrm{d}_{r^2}^{2}\square+\frac{2}{r} \mathrm{d}_{r}\square-\frac{\ell(\ell+1)}{r^{2}}\square,
\end{gather}
and the coupling coefficients, which all depend on $m$, are given by
\begin{gather}
    \alpha_{\ell-1}^{\ell}=\alpha_{\ell}^{\ell-1}=\sqrt{\frac{\ell^{2}-m^{2}}{(2 \ell-1)(2 \ell+1)}},\\
    \beta_{\ell-1}^{\ell}=(\ell-1) \alpha_{\ell-1}^{\ell}, \quad \beta_{\ell+1}^{\ell}=-(\ell+2) \alpha_{\ell+1}^{\ell},\\
    \gamma_{\ell-1}^{\ell}=\frac{\ell-1}{\ell} \alpha_{\ell-1}^{\ell}, \quad
    \gamma_{\ell+1}^{\ell}=\frac{\ell+2}{\ell+1} \alpha_{\ell+1}^{\ell}.
\end{gather}
Since we only consider the dominant quadrupolar tidal component $\ell = m = 2$, the forcing term can be written as
\begin{equation}
    \vec{f}^* = f^{\ell,m}_{\rm R}(r)\vec{R}_{\ell}^m + f^{\ell,m}_{\rm S}(r)\vec{S}_{\ell}^m + f^{\ell,m}_{\rm T}(r)\vec{T}_{\ell}^m,
\end{equation}
with 
\begin{equation}
    f^{\ell,m}_{\rm R}= \left[\tilde{\omega}^2 \xi_{r, \ell}^{\rm nw} + \frac{m\tilde{\omega}}{\ell(\ell+1)}\left(2\xi_{r, \ell}^{\rm nw} + r \mathrm{d}_r \xi_{r, \ell}^{\rm nw}\right)\right]\delta(\ell-2)\delta(m-2),
\end{equation}
\begin{multline}
    f^{\ell,m}_{\rm S} = \left[\frac{m\tilde{\omega}}{\ell^2(\ell+1)^2} \left(\left(\ell(\ell+1)+2\right)\xi_{r, \ell}^{\rm nw} + r \mathrm{d}_r \xi_{r, \ell}^{\rm nw}\right) +\right.\\
    \left.\frac{\tilde{\omega}^2}{\ell(\ell+1)} \left(2\xi_{r, \ell}^{\rm nw} + r \mathrm{d}_r \xi_{r, \ell}^{\rm nw}\right) \right] \delta(\ell-2)\delta(m-2),
\end{multline}
\begin{equation}
    f^{\ell,m}_{\rm T}=-i\tilde{\omega} \frac{\alpha_{\ell-1}^{\ell}}{\ell^2} \left(-(\ell-2)\xi_{r, \ell-1}^{\rm nw} + r \mathrm{d}_r \xi_{r, \ell-1}^{\rm nw}\right)\delta(\ell-3)\delta(m-2),
\end{equation}
obtained by projecting Eq.\,\eqref{eq:foricng} on the spherical harmonics basis.

\subsubsection{Boundary conditions}\label{subsubsect:bc}
Given our specific emphasis on (gravito-)inertial modes while excluding surface gravity modes, we can adopt the classical boundary conditions established in pioneering studies by \cite{Dintrans1999, Dintrans2000, Valdettaro2007, Ogilvie2004,Ogilvie2005,Ogilvie2007,Ogilvie2009, Rieutord2010}.
Namely, we employ impenetrable and stress-free boundary conditions, while assuming that the spheres bounding the fluid domain can absorb any flux of heat or chemical elements while remaining at constant temperature and molecular weight. Namely, the radial functions must satisfy the following inner ($r=\eta$) and outer ($r=1$) boundary conditions
\begin{gather}
         u^{m}_{\ell} = 0,\\
         \frac{\mathrm{d} v^{m}_{\ell}}{\mathrm{d} r} + \frac{u^m_{\ell}-v^m_{\ell}}{r}=0,\\
         \frac{\mathrm{d}}{\mathrm{d} r}\left(\frac{w^{m}_{\ell}}{r}\right)=0, \\ t^{m}_{\ell} =0, \\
         \mu^{m}_{\ell} =0.
\end{gather}

\section{Results}\label{sec:5}
Having established the framework for our numerical work, we now present the numerical results. We will discuss the basic properties of the forced waves before considering how the dissipative properties depend on the model’s key parameters, and discuss the implications for astrophysical tidal evolution.
Our calculations are focused on the frequency range of $-1<\tilde{\omega}<-0.5$, which is directly relevant to the tidal frequencies of the Galilean moons. It is worth noting that the negative tidal frequency indicates that the tidal forcing is retrograde in the co-rotating frame with the planet, based on our convention.

\subsection{Energies}
\begin{figure*}
    \centering
\begin{minipage}{0.33\textwidth}
\centering
\resizebox{\hsize}{!}{\includegraphics{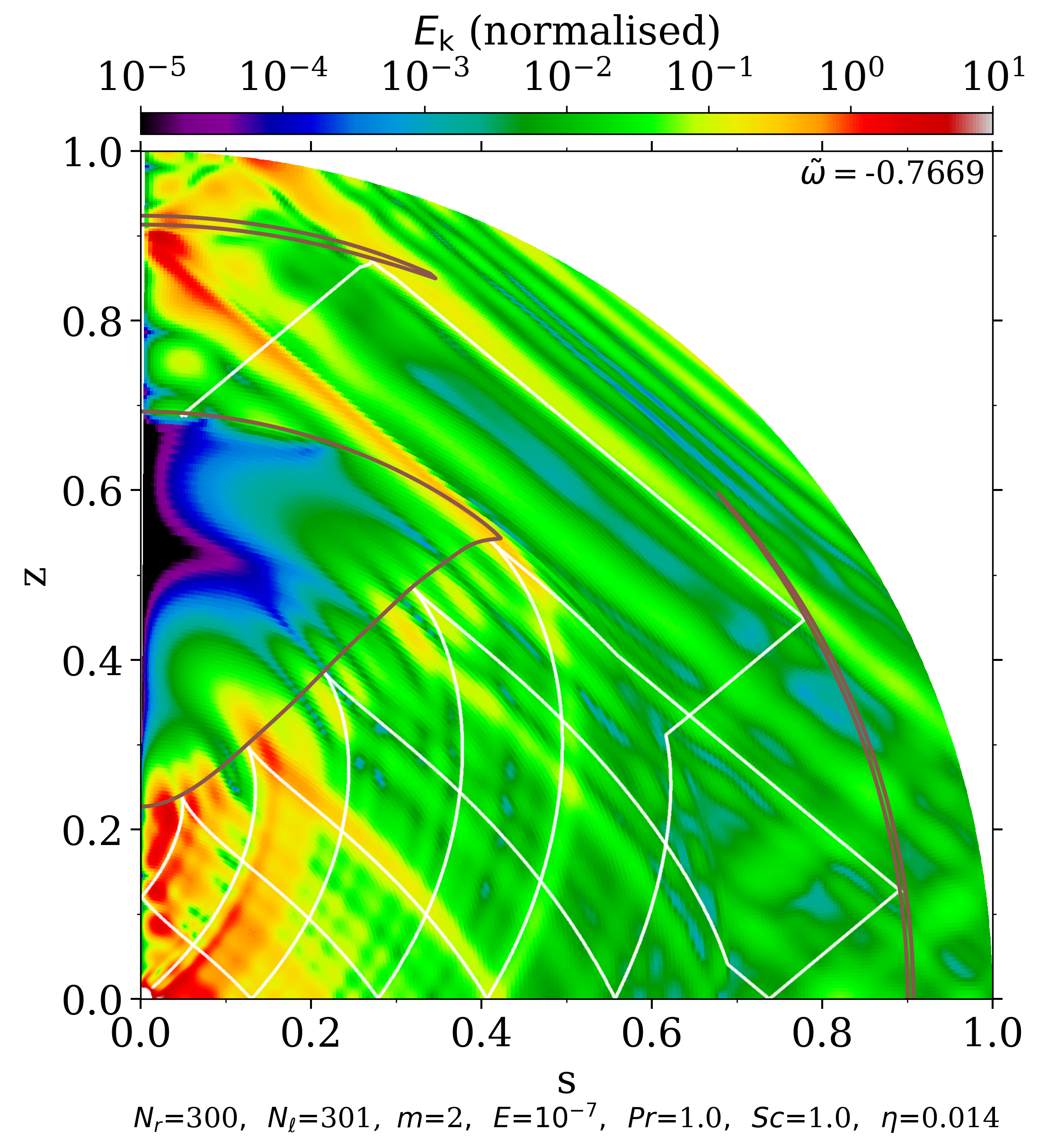}}
\end{minipage}
\begin{minipage}{0.33\textwidth}
\centering
\resizebox{\hsize}{!}{\includegraphics{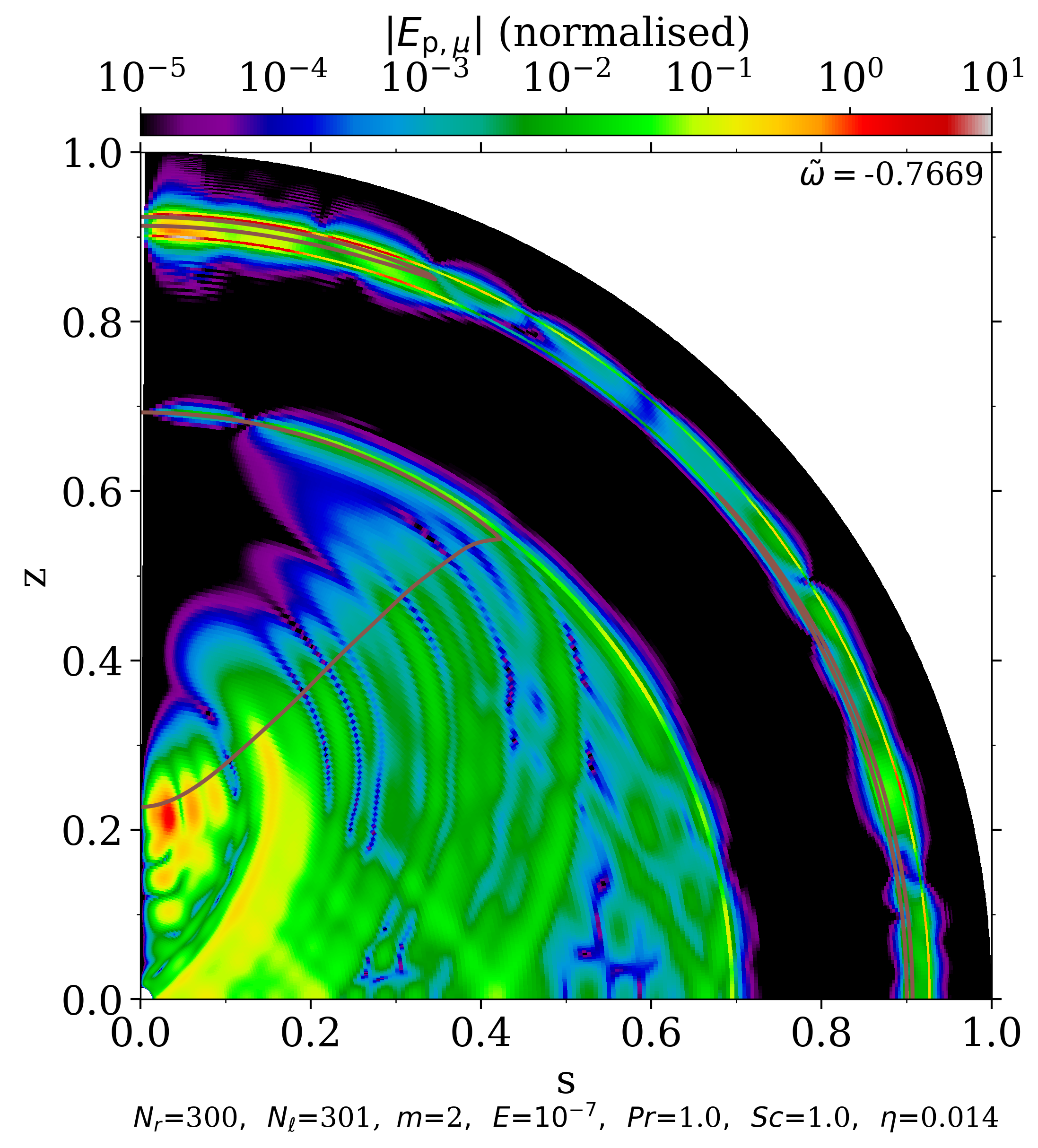}}
\end{minipage}
\begin{minipage}{0.33\textwidth}
\centering
\resizebox{\hsize}{!}{\includegraphics{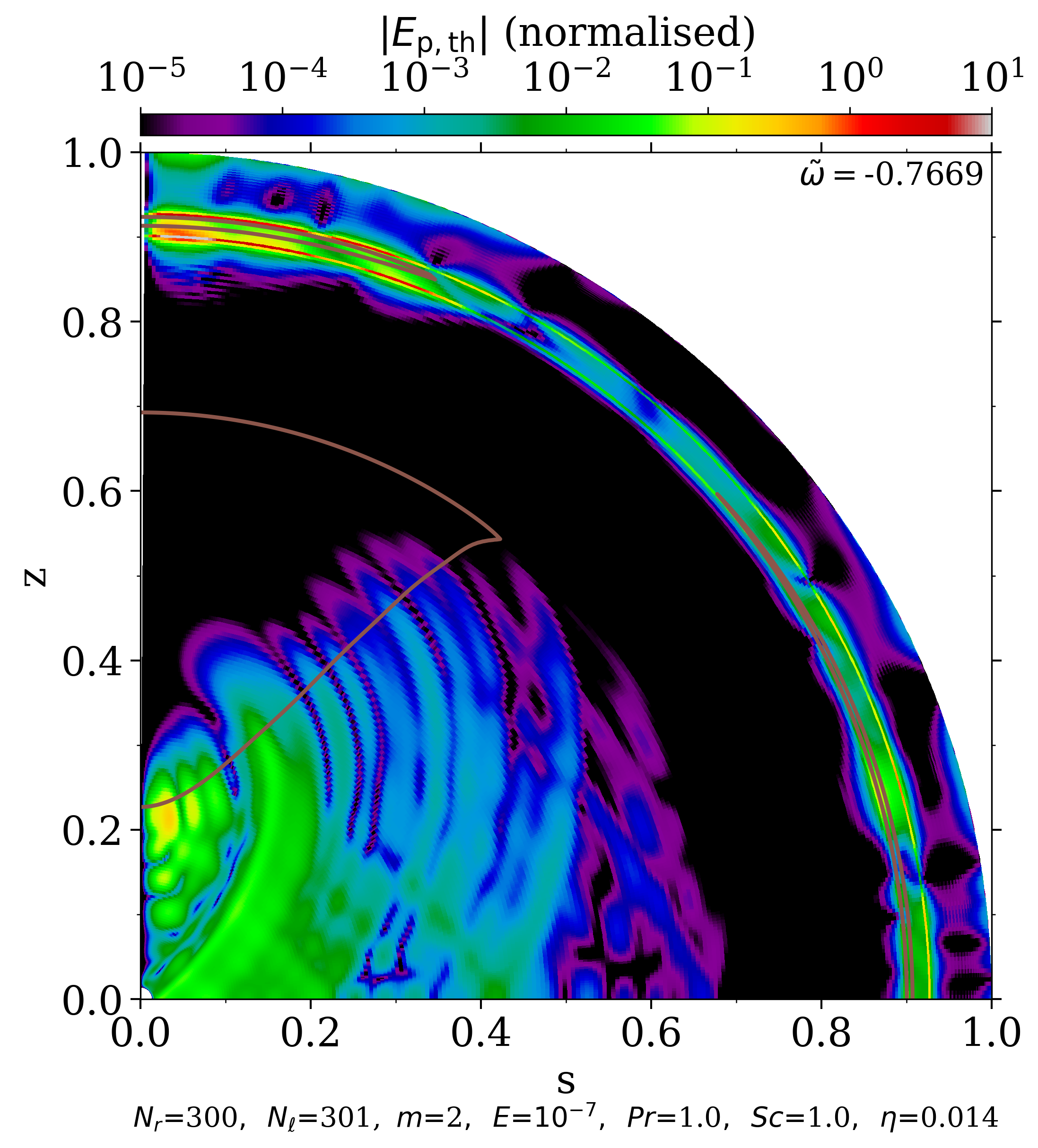}}
\end{minipage}
    \caption{Meridional cut ($z= r \cos \theta$ and $s=r \sin \theta$) of the dimensionless kinetic energy (left), potential energy associated with chemical (middle), and  potential energy associated thermal  stratification (right) of the forced mode $\tilde{\omega}=-0.76$ for $m=2$, $\mathrm{E}=10^{-7}$, $\mathrm{Pr}=\mathrm{Sc}=1$ with a spatial resolution $(N_r,\,N_\ell)=(300,\,301$). The trajectories of characteristics are represented with white curves, while the surfaces at which they undergo reflections with brown curves.}
    \label{fig:energy}
\end{figure*}

\begin{figure}
\centering
\resizebox{\hsize}{!}{\includegraphics{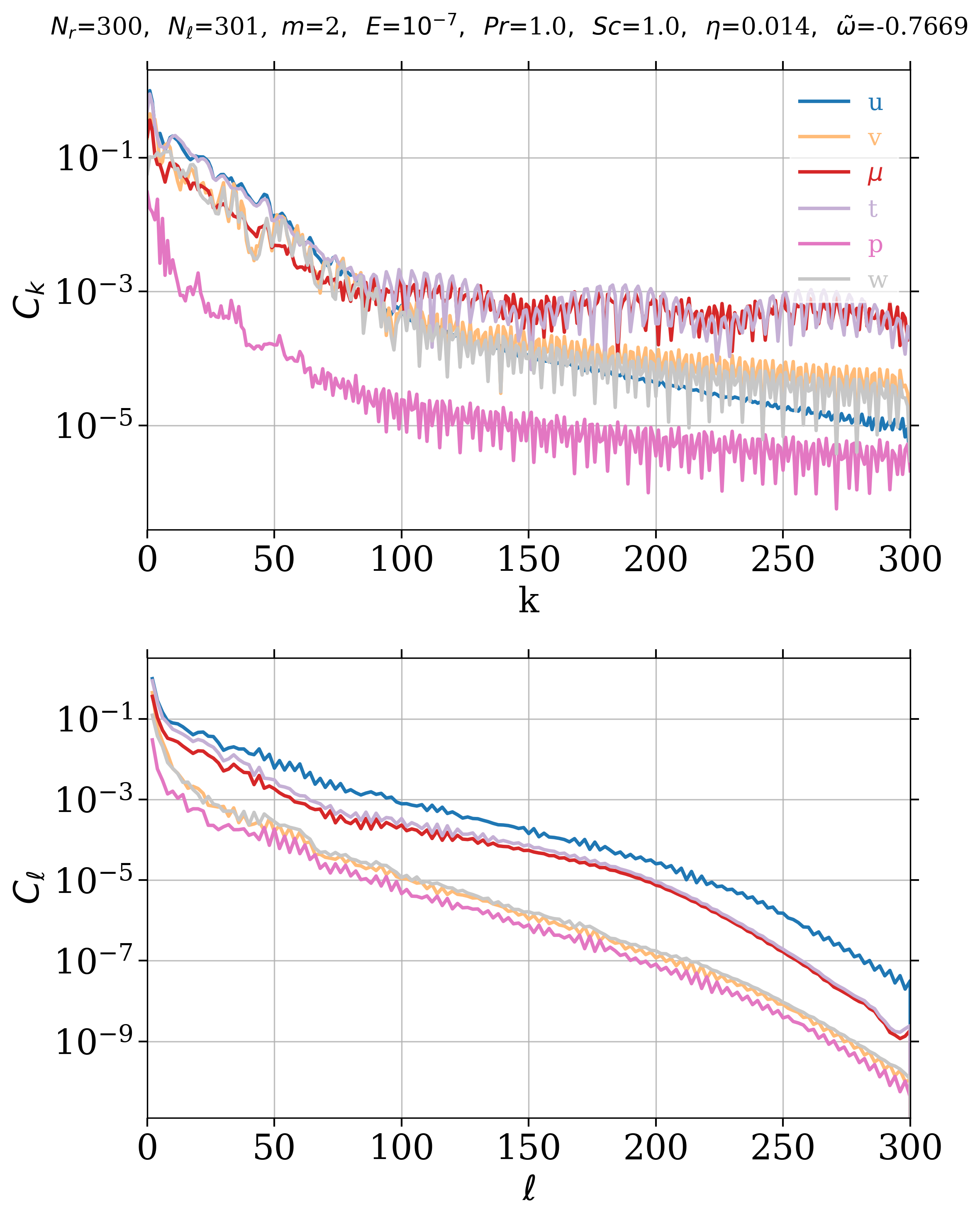}}
\caption{Spectral content of the velocity field components $(\mathrm{u, v, w})$, pressure $(p)$, temperature $(t)$, and molecular weight $(\mu)$ of the forced mode $\tilde{\omega}=-0.76$ for $m=2$, $\mathrm{E}=10^{-7}$, $\mathrm{Pr}=\mathrm{Sc}=1$. Chebyshev and spherical harmonics coefficients are shown in the top and bottom panels, respectively.}
\label{fig:spectral}
\end{figure}

In this section, we present our numerical results for the different types of energy examined in our study, with a specific focus on the forced mode $\tilde{\omega}=-0.76$ (the frequency exited in Jupiter by Io for $m=n=2$). We adopt typical values for various dimensionless numbers, namely $\mathrm{E}=10^{-7}$, $\mathrm{Pr}=\mathrm{Sc}=1$. Furthermore, we set the aspect ratio to $\eta=0.014$ (because of the solid core of size $1.4\%$ of radius, see Sec.\,\ref{subsec:struc_model}).
The left panel of Fig.\,\ref{fig:energy} illustrates the spatial distribution of the kinetic energy of this mode in a meridional quarter-plane since it is symmetrical with respect to the equator. Notably, we can observe two distinct types of modes. Firstly, there are gravito-inertial modes that exist within the inner stably stratified regions (and theoretically within the thin outer stably stratified layer, but it is very thin so it is not clear here). As expected, we find equatorial trapping of sub-inertial ($\tilde{\omega}<1$) gravito-inertial modes \citep[e.g.][]{Dintrans1999, Dintrans2000, Mathis2009}.
Secondly, we have inertial modes present in the two convective zones, which are separated by the thin stably stratified layer. These modes exhibit multiple reflections at the boundaries of their propagation zones, following specific trajectories known as attractors \citep{Maas1995}. We note that the attractor starting from the critical latitude \citep[e.g.][]{Rieutord2001, Rieutord2018} in the inner convective zone and reflected at the pole, at the surface, at the equator and at the interface with the innermost stably-stratified zone seems
to appear as well, regardless of the thin intermediate stably stratified region.

We represent also in this figure the trajectories of characteristics with white curves, while the surfaces at which they undergo reflections are depicted by brown curves. These paths of characteristics are calculated based on the second-order partial differential equation satisfied by the pressure perturbation in the inviscid and short-wavelength approximations (see \citet{Mirouh2016} for the detailed derivation). The inclusion of these curves provides a valuable means of understanding the solutions to non-dissipative problems and validation of numerical calculations.
We find that the patterns formed by the characteristics are in very good agreement with the numerical calculation, especially in the inner stably stratified layers. In convective zones, the paths of characteristics follow straight lines that maintain a constant angle relative to the rotation axis (z-axis) in order to respect the inertial wave dispersion relation. In contrast, stably stratified regions introduce a distinct behaviour where the characteristics become curved, owing to the distortion caused by the presence of the stable stratification. 

The middle and right panels of Fig.\,\ref{fig:energy} reveal that the chemical and thermal energies primarily concentrate within the stably stratified regions, as the thermal and chemical Brunt-Väisälä frequencies approach zero within convective zones. Consequently, at the interfaces of the convective zones, both potential energies have a finite transition to zero.

To ensure the numerical convergence for this mode, we employ a spatial resolution of $(N_r,\, N_\ell)=(300,\,301)$. This convergence can be appreciated by inspection at Fig.\,\ref{fig:spectral}, where we display the spectral content of the velocity field components, pressure, temperature, and molecular weight for the same forced mode. In the top panel, we show the maximum Chebyshev coefficients $C_k$ as a function of the Chebyshev order $k$, selecting the highest value among all the spherical harmonics coefficients corresponding to a given $k$. Likewise, the bottom panel displays the maximum spherical harmonics coefficients $C_\ell$ as a function of the spherical harmonic degree $\ell$, considering the highest value among all Chebyshev coefficients.
This spatial resolution $(N_r,\, N_\ell)=(300,\,301)$  has proved to be sufficient up to values of $E \approx 10^{-8}$.

\subsection{Dissipation spectra}
\begin{figure*}
    \centering
    \resizebox{\hsize}{!}{\includegraphics{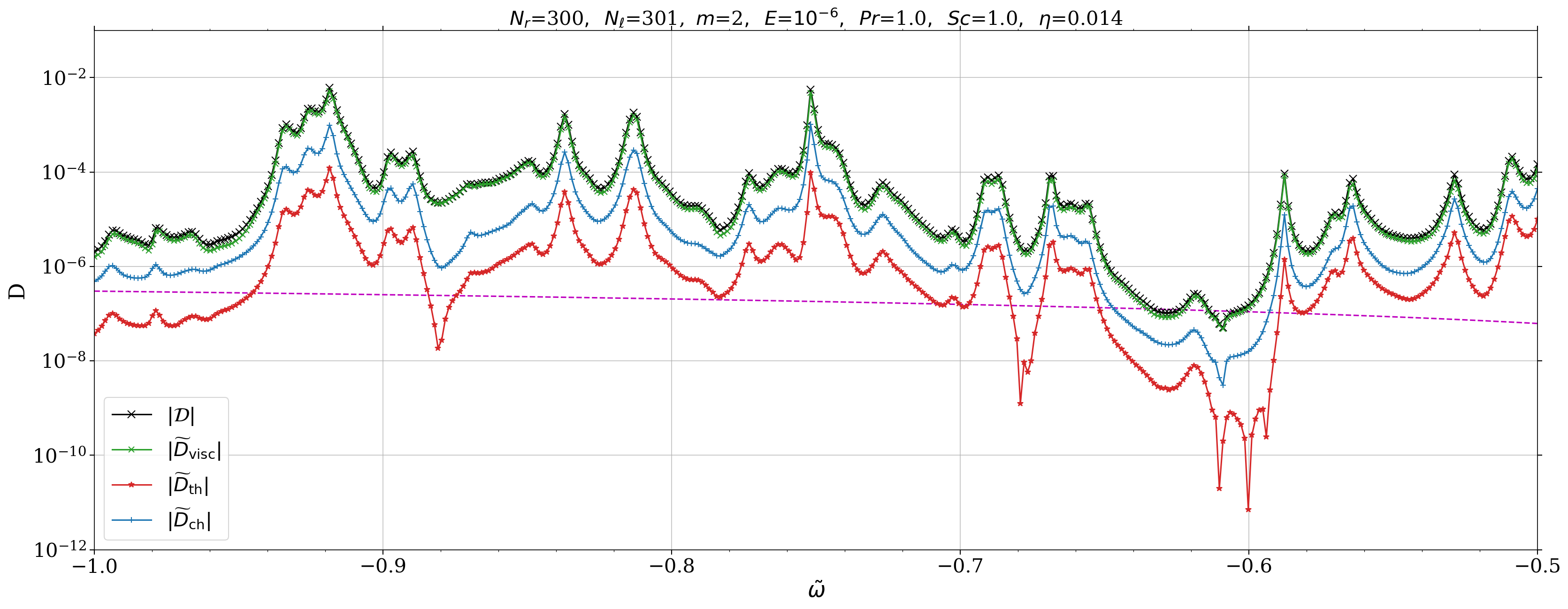}}
    \resizebox{\hsize}{!}{\includegraphics{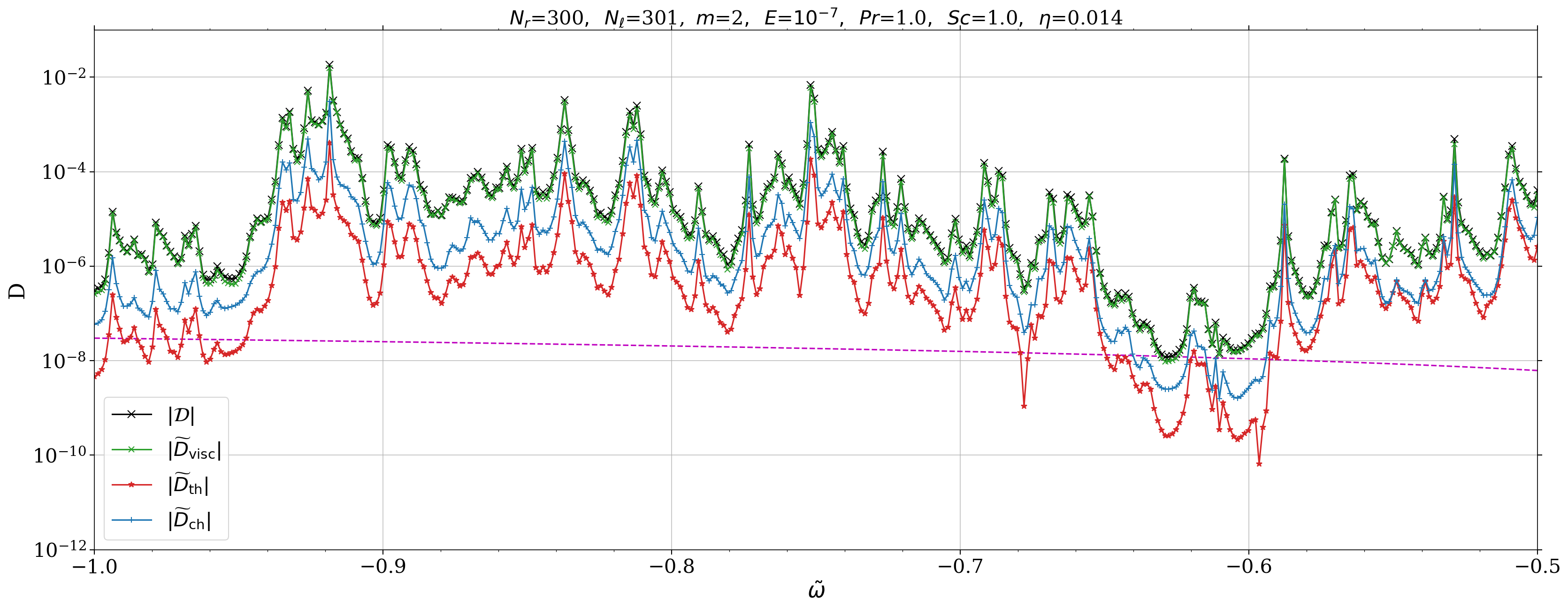}}
    \resizebox{\hsize}{!}{\includegraphics{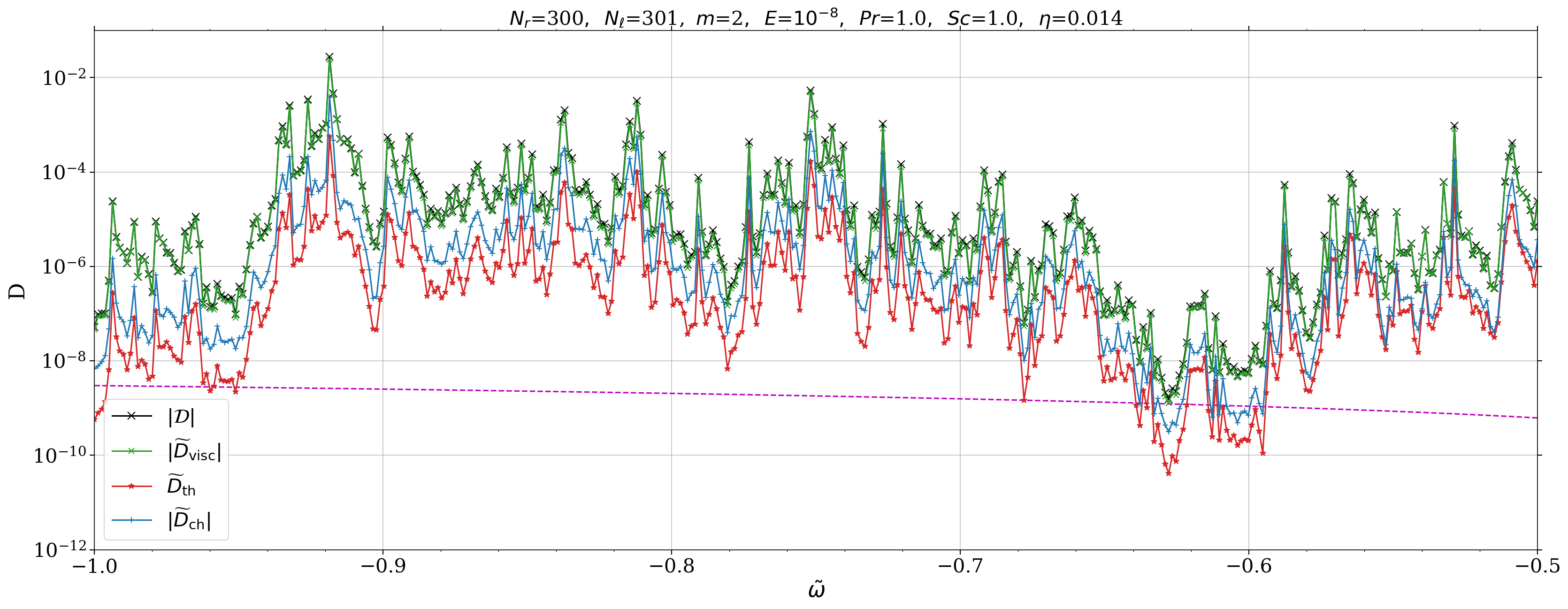}}
    \caption{Dissipation spectra for $m=2$, $\mathrm{Pr}=\mathrm{Sc}=1$, and $\mathrm{E}=10^{-6}$ (Top), $\mathrm{E}=10^{-7}$ (middle) and $\mathrm{E}=10^{-8}$ (Bottom) with a spatial resolution $(N_r,\,N_\ell)=(300,\,301$). The total dissipation is represented in black crosses, and its viscous, thermal, and molecular  contributions are represented in green points, red stars, and  blue plus signs, respectively. The magenta dashed line indicate the viscous (total) dissipation in the case of purely convective interior ($N_\mu^2=N_{\rm t}^2=0$).}
    \label{fig:diss_spectra}
\end{figure*}

We analyse three distinct forms of dissipation: viscous, thermal, and chemical. The total dissipation is defined as follows:
\begin{equation}
    \mathcal{D} (\tilde{\omega}) = \widetilde{D}_{\mathrm{th}} + \widetilde{D}_{\mathrm{ch}} + \widetilde{D}_{\mathrm{visc}}.
\end{equation}
Fig.\,\ref{fig:diss_spectra} shows the viscous ($\widetilde{D}_{\mathrm{visc}}$), thermal ($\widetilde{D}_{\mathrm{th}}$), molecular ($\widetilde{D}_{\mathrm{ch}}$) and total ($\mathcal{D}$) dissipation rates integrated over the volume as a function of the normalised forcing frequency ($\tilde{\omega}$) for $m=2$, $\mathrm{E}=\{10^{-6}, 10^{-7}, 10^{-8}$\}, and $\mathrm{Pr}=\mathrm{Sc}=1$. 
We observe a significant frequency dependence, indicating a strong relationship between dissipation and the forcing frequency. Moreover, our analysis reveals that the dominant mechanism contributing to dissipation is viscosity, surpassing both thermal and chemical dissipations in magnitude.
We ensure that the total energy is conserved $\mathcal{D}\approx \widetilde{P}_{\rm tide}$ to a given degree of confidence (maximum relative error of $5\%$). Note that given our boundary conditions (Sec.\;\ref{subsubsect:bc}), $P_{\rm acou}\approx 0$.

Following the comparison method adopted by \citet{Andre2019} in Cartesian coordinates, we also computed the dissipation spectra for the old vision of Jupiter's interior, where there is a single purely convective zone extending from $r=\eta=0.014$ to $r=1$. In this scenario, the only form of dissipation present is viscous dissipation, as thermal and chemical dissipation are negligible due to $N_\mu^2=N_{\rm t}^2=0$. The dissipation due to viscosity is represented by the magenta dashed line in Fig.\,\ref{fig:diss_spectra}.  We can see that the spectra in this case exhibit a smooth profile, devoid of any pronounced peaks at specific frequencies, unlike the four-layer model. Additionally, it is worth noting that the dissipation in this case is significantly weaker, ranging from two to four orders of magnitude lower.

\begin{figure*}
    \centering
    \resizebox{\hsize}{!}{\includegraphics{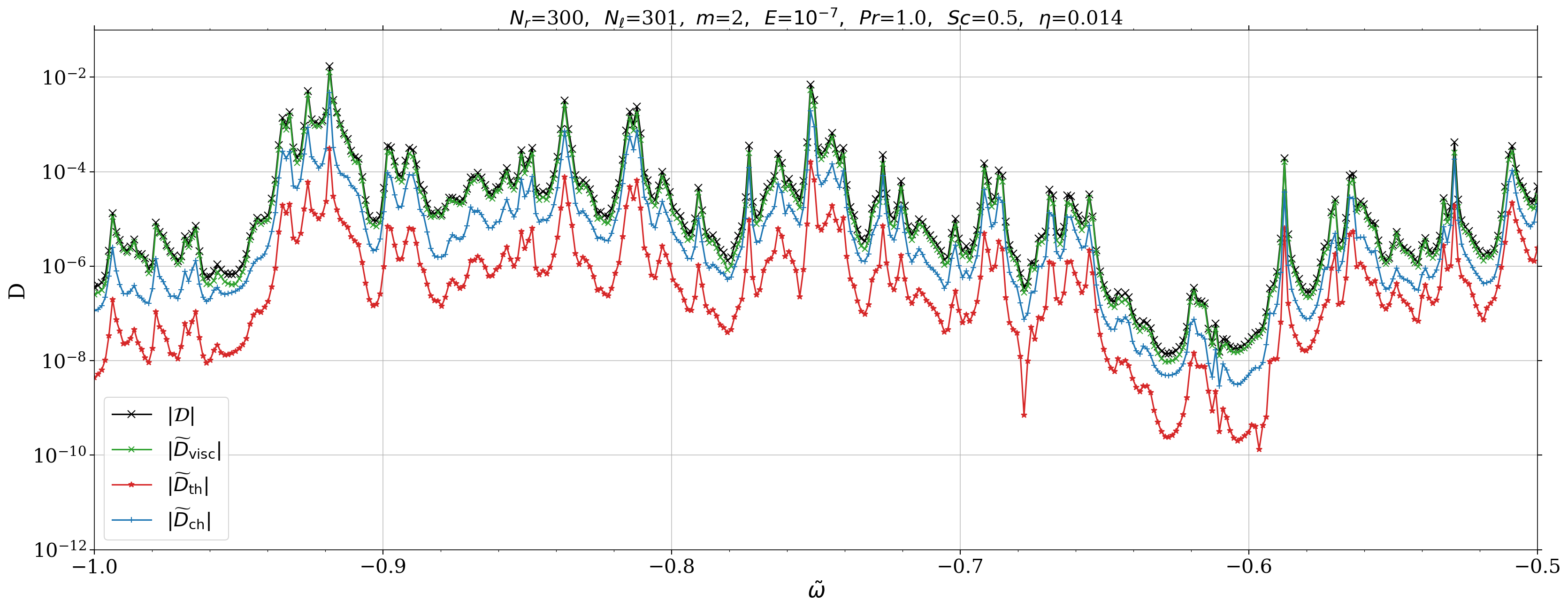}}
    \resizebox{\hsize}{!}{\includegraphics{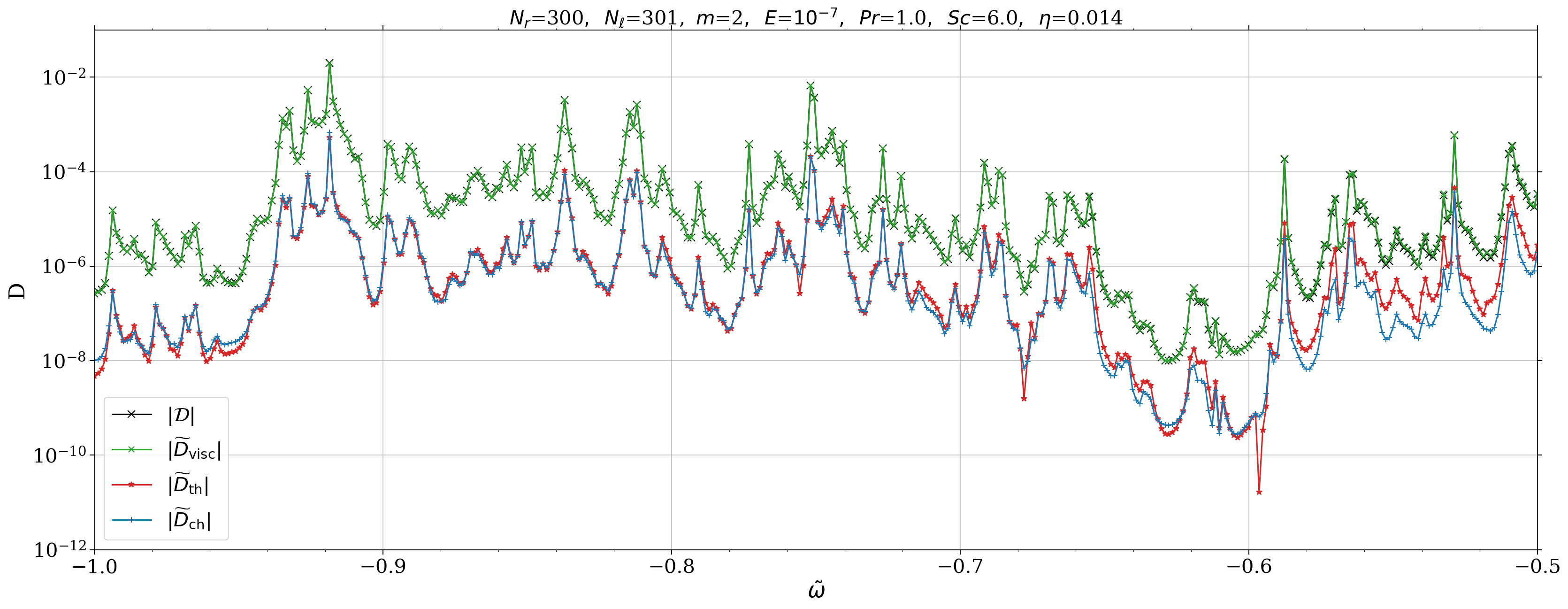}}
    \caption{Same as the middle panel of Fig.\,\ref{fig:diss_spectra} but for  $\mathrm{Sc}=0.5$ (Top) and $\mathrm{Sc}=6$ (Bottom).}
    \label{fig:spec_sc}
\end{figure*}

\begin{figure*}
    \centering
    \resizebox{\hsize}{!}{\includegraphics{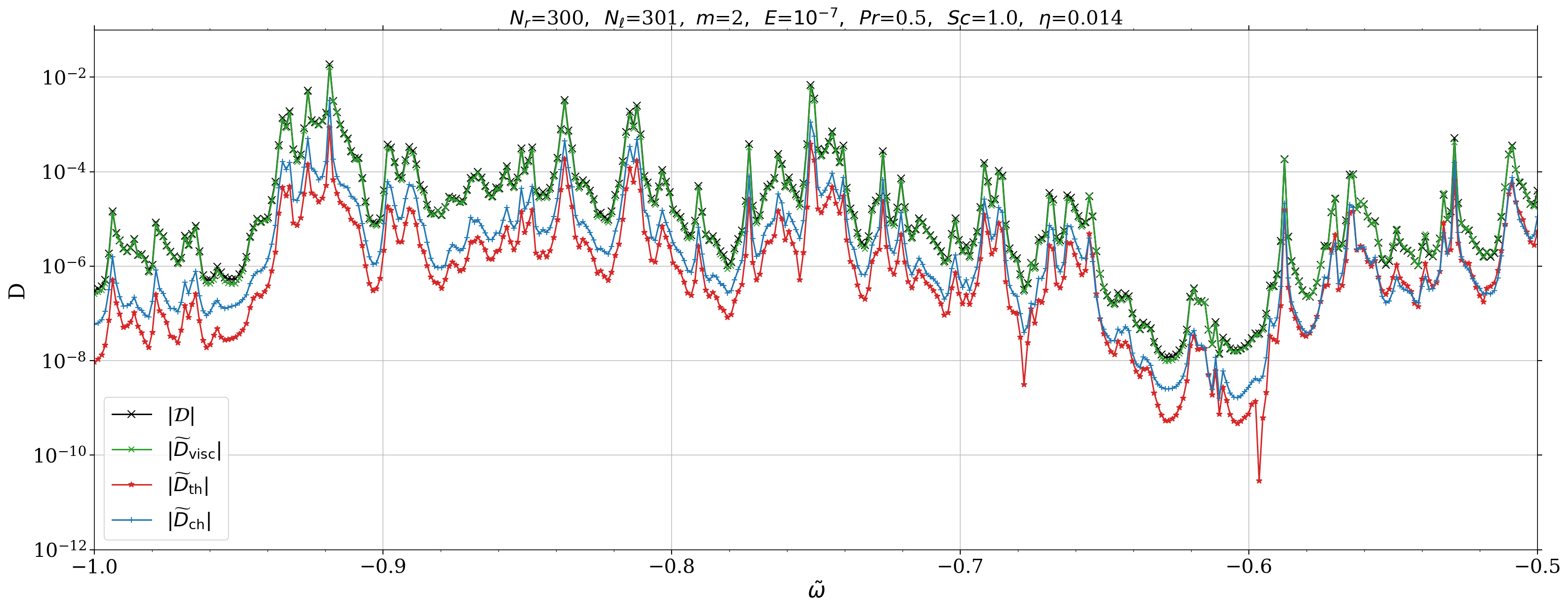}}
    \resizebox{\hsize}{!}{\includegraphics{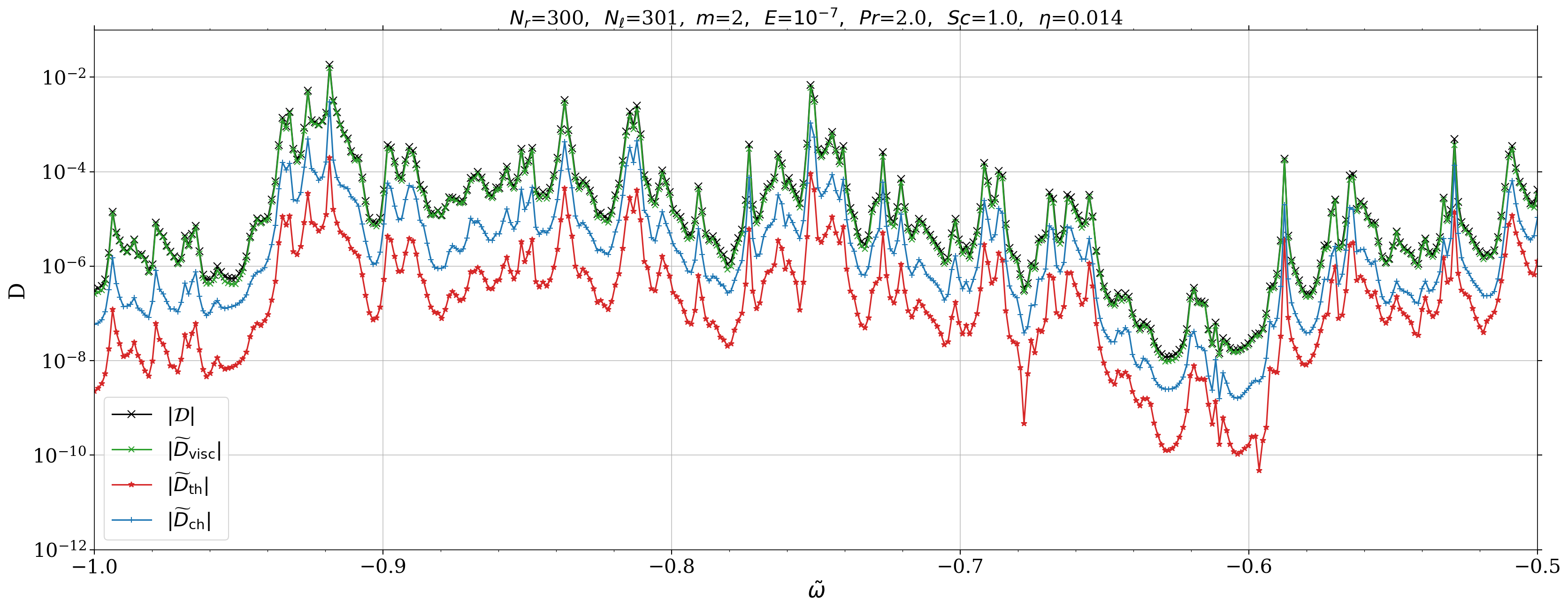}}
    \caption{Same as the middle panel of Fig.\,\ref{fig:diss_spectra} but for  $\mathrm{Pr}=0.5$ (Top) and $\mathrm{Pr}=2$ (Bottom).}
    \label{fig:spec_pr}
\end{figure*}

We focus also on the influence of Ekman number variations on dissipation spectra.
Our results reveal that varying the Ekman number has a significant impact on the energy dissipation. Indeed, we find an increase in the number of peaks in the dissipation spectra as the Ekman number decreases.
More specifically, the decrease in the Ekman number leads to lower viscosity which results in higher and narrower resonance peaks associated with gravito-inertial modes, making the spectrum more complex, whereas all peaks are smoothed when a higher viscosity is used.
This result is consistent with the predictions of \citet{Auclair2015} who studied in a Cartesian box the dissipation of gravito-inertial waves by viscosity and thermal diffusion in a stably stratified medium.
We also explore the impact of the Schmidt ($\mathrm{Sc} = \{0.5, 6\}$) and Prandtl ($\mathrm{Pr} = \{0.5, 2\}$) numbers in Figs. \ref{fig:spec_sc} and \ref{fig:spec_pr}, respectively. 
We find that decreasing the Schmidt (Prandtl) number increases the molecular (thermal) dissipation. However, the total dissipation is not modified, since the viscous dissipation is dominant in this parameter regime.

\subsection{Quality factor and Love number}
To establish a connection between our numerical computations and observations, it is necessary to calculate the imaginary component of the Love number from the total dissipation. This calculation enables us to conduct a comprehensive analysis by comparing our numerical models with actual observations by performing quantitative and qualitative comparisons. 
\begin{figure*}
    \centering
    \resizebox{\hsize}{!}{\includegraphics{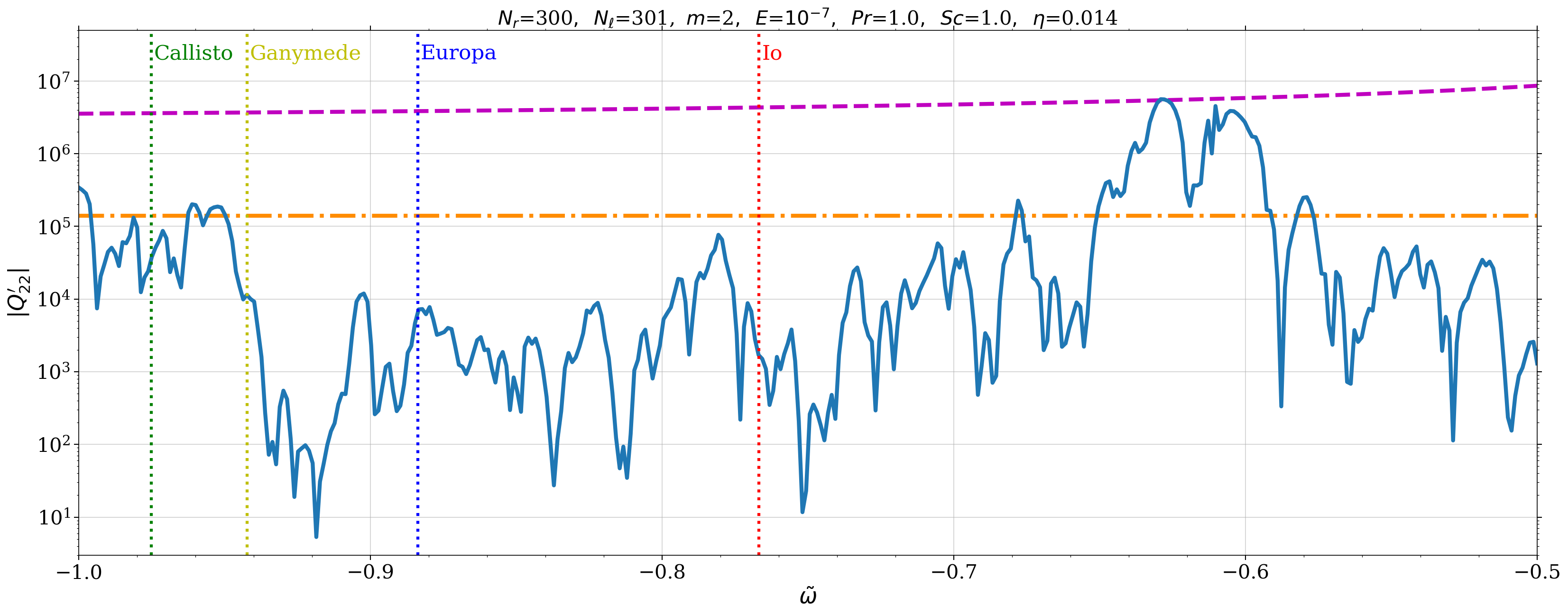}}
    \caption{Modified tidal quality factor as a function of the tidal frequency for $m=2$, $\mathrm{E}=10^{-7}$, and $\mathrm{Pr}=\mathrm{Sc}=1$.  Vertical dotted lines indicate the tidal frequencies for the four Galilean Moons of Jupiter (from right to left: Io, Europa, Ganymede, Callisto). The magenta dashed line indicates the values of these quantities in the case of a purely convective interior ($N_\mu^2=N_{\rm t}^2=0$). The dash-dotted orange line marks the observed value of this quantity due to Io. }
    \label{fig:Q}
\end{figure*}

\begin{figure*}
    \centering
    \resizebox{\hsize}{!}{\includegraphics{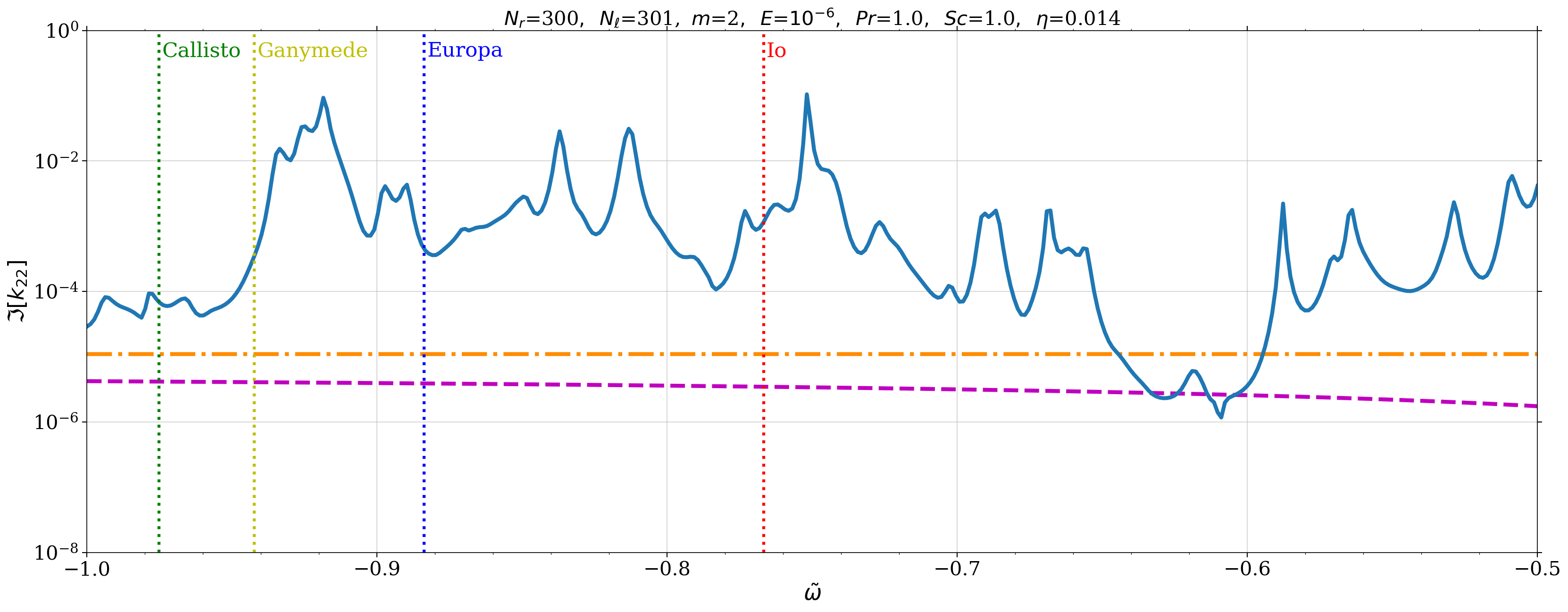}}
    \resizebox{\hsize}{!}{\includegraphics{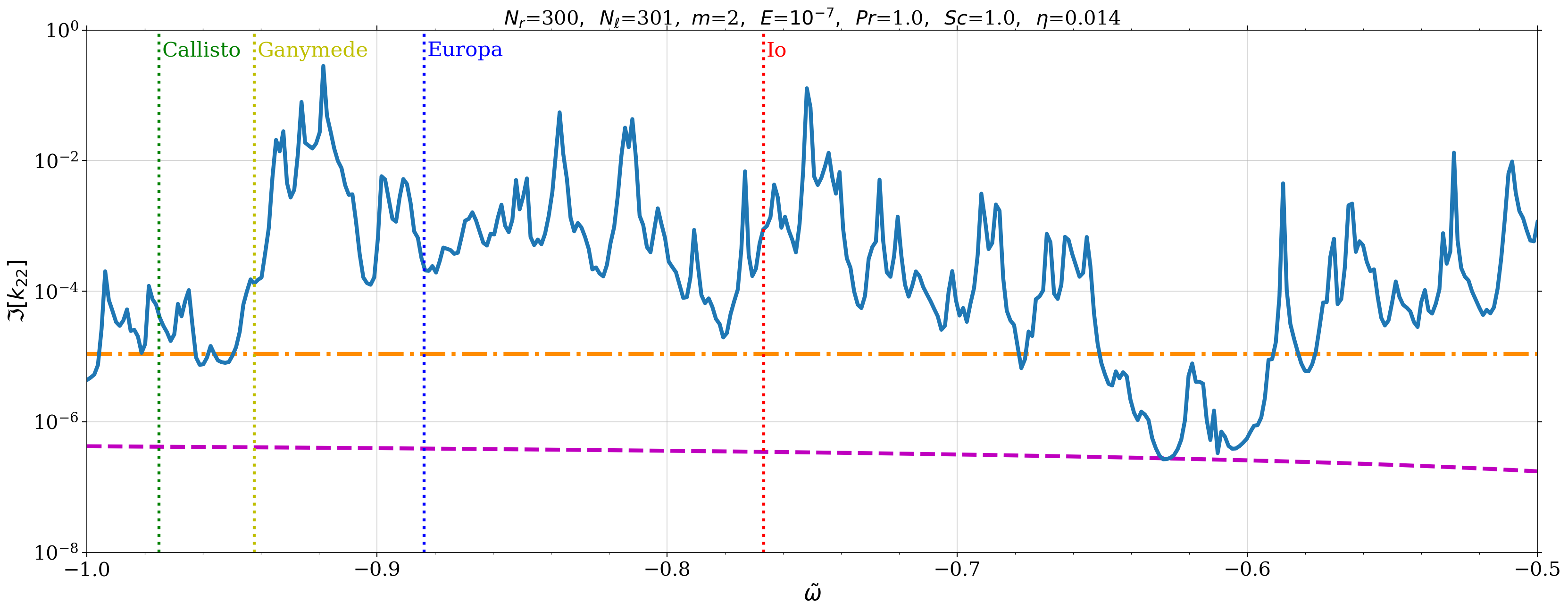}}
    \resizebox{\hsize}{!}{\includegraphics{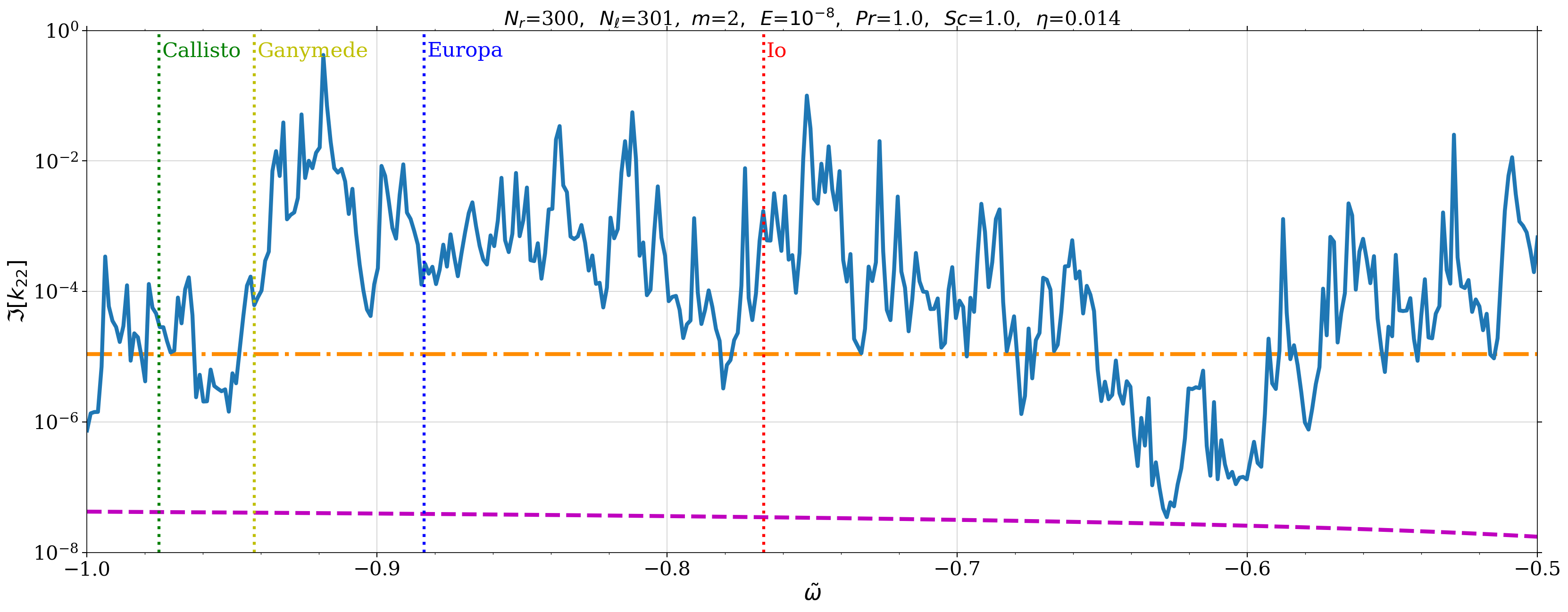}}
    \caption{Imaginary part of the Love number as a function of the tidal frequency for $m=2$, $\mathrm{Pr}=\mathrm{Sc}=1$, and $\mathrm{E}=10^{-6}$ (Top), $\mathrm{E}=10^{-7}$ (middle) and $\mathrm{E}=10^{-8}$ (Bottom).  Vertical dotted lines indicate the tidal frequencies for the four Galilean Moons of Jupiter (from right to left: Io, Europa, Ganymede, Callisto). The magenta dashed line indicates the values of these quantities in the case of a purely convective interior ($N_\mu^2=N_{\rm t}^2=0$). The dash-dotted orange line marks the observed value of this quantity due to Io. }
    \label{fig:Imk}
\end{figure*}

The imaginary part of the Love number $\Im\left[k_{\ell m}\right]$ plays a crucial role in characterising the response of a celestial body to tidal forces, capturing the phase difference between the applied tidal forcing and the resulting response. It represents the transfer of energy and angular momentum within the system. 
By establishing a relation between the overall dissipation and the imaginary part of the Love number, we gain valuable insights into the evolution of the system. This relation can be expressed as in \cite{Ogilvie2013}
\begin{equation}\label{eq:imk}
    \Im\left[k_{\ell m}(\tilde{\omega})\right] = \frac{G M}{4R^3\Omega^2} \frac{8 \pi}{(2 \ell+1)\tilde{\omega}} \mathcal{D},
\end{equation}
then we can define the modified tidal quality factor as
\begin{equation}\label{eq:Q}
    {Q^\prime}_{\ell m}(\tilde{\omega}) = \text{sign}(\tilde{\omega}) \frac{3}{2 \Im\left[k_{\ell m}(\tilde{\omega})\right]},
\end{equation}
which has the advantage of combining the tidal quality factor $Q$ with the real part of the Love number $\Re[k_{\ell m}]$
\begin{equation}
    \Im\left[k_{\ell m}(\tilde{\omega})\right]= \text{sign}(\tilde{\omega}) \frac{\Re[k_{\ell m}(\tilde{\omega})]}{Q_{\ell m}(\tilde{\omega})}.
\end{equation}
\cite{Lainey2009} have fitted a dynamical model including parameterised tidal dissipation, to astrometric observations from 1891 to 2007
of the Galilean satellites. They found that $\Im\left[k_{2 2}\right] = 1.1 \times 10^{-5}$ ($Q_{2 2}^\prime = -1.4 \times 10^5$), for the asynchronous tide ($\ell=m=n=2$) due to Io ($\tilde{\omega}=-0.76$).

Using Eqs.\,\eqref{eq:imk}\,\&\,\eqref{eq:Q} we compute  the imaginary part of the Love number and the modified tidal quality factor, and we represent them as a function of the normalized forcing frequency in Fig.\,\ref{fig:Q} and the middle panel of Fig.\,\ref{fig:Imk} for $m=2$, $\mathrm{E}=10^{-7}$, and $\mathrm{Pr}=\mathrm{Sc}=1$.
We find a significant discrepancy between computed values of the imaginary part of the Love number (the modified tidal quality factor) due to Io and the observed ones, differing by roughly two order of magnitude (the computed imaginary part of the Love number is $8.8 \times 10^{-4}$ whereas the observed one is $1.1 \times 10^{-5}$). Consequently, our calculations tend to overestimate the amplitude of tidal dissipation. Conversely, when examining the purely convective model, we observe an underestimation of tidal dissipation by approximately one and a half orders of magnitude.

We investigate the influence of varying Ekman number on the imaginary part of the Love number. The outcomes are illustrated in Fig.\,\ref{fig:Imk} for $\mathrm{E}=\{10^{-6}, 10^{-7}, 10^{-8}$\}. We find that decreasing the Ekman number impacts the  imaginary part of the Love number (the dissipation) by increasing the number of peaks. But for the frequency associated with the forcing imposed by Io, the impact is small because we are not on a resonance (a peak), and the imaginary part of the Love number does not vary significantly. We find the same result for the Schmidt and Prandtl numbers. Their impact on the total dissipation is very weak, therefore they do not influence the imaginary part of the Love number.

Eventually, we find that stable stratification plays a crucial role in explaining the high dissipation. This conclusion was also highlighted by \citet{Andre2019}, who investigated tidal dissipation in a rotating semi-convective region with a Cartesian box model.
In addition, \citet{Lin2023} and \citet{Dewberry2023} have also studied tidal responses in some simplified scenarios conceivable for Jupiter's interior with stably stratified layers, taking only the viscous diffusion into account. 
Our results confirm their results with taking into account the three possible diffusion mechanisms, which are dominated by the viscous one, and more realistic internal structure models for Jupiter.

\subsection{Impact of the external stably stratified layer: four zones vs two zones models}
\begin{figure}
    \centering
    \resizebox{\hsize}{!}{\includegraphics{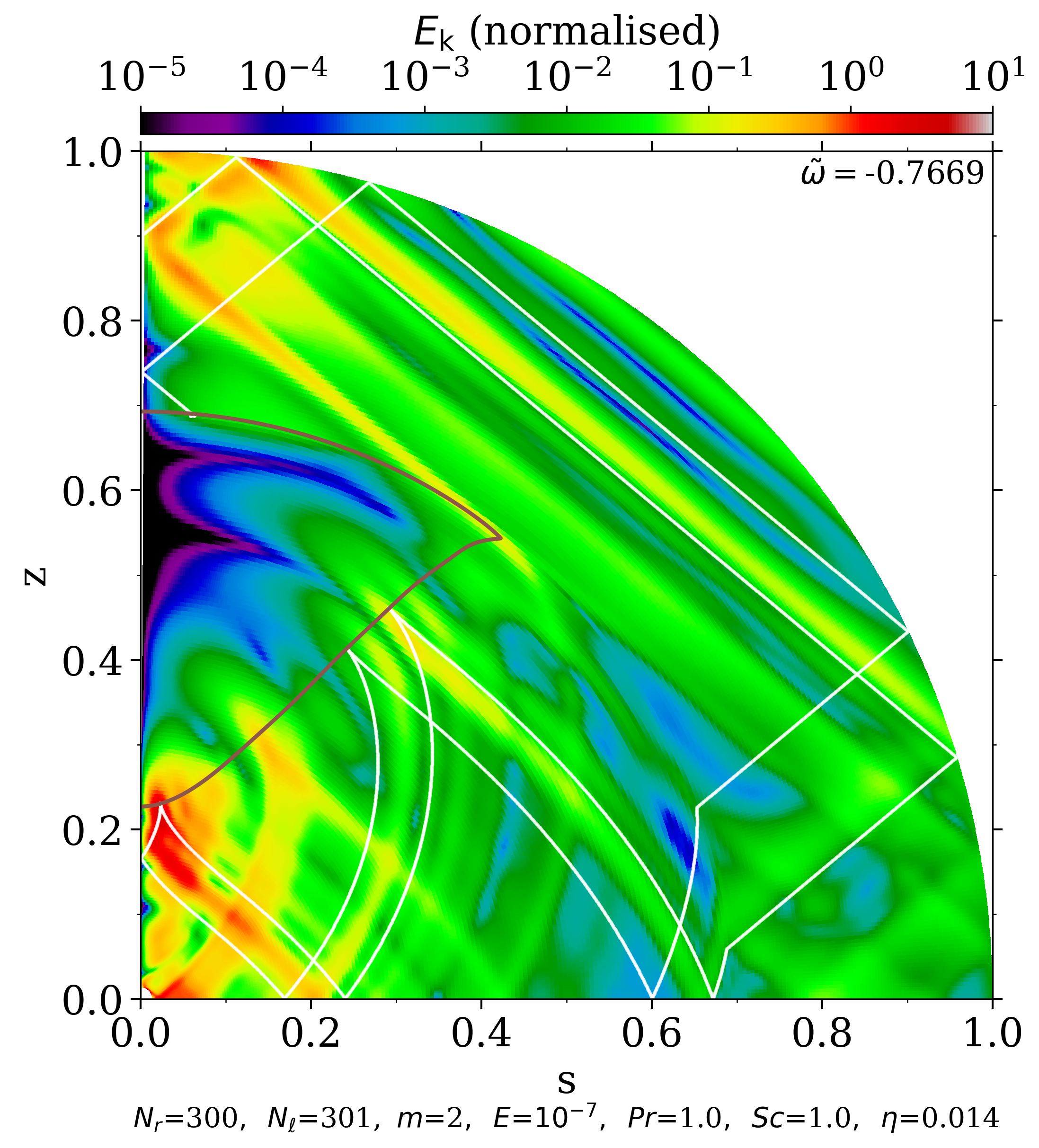}}
    \caption{Same as the left panel of Fig.\,\ref{fig:energy}, but for an interior model without the thin external stably stratified layer (two zones model).}
    \label{fig:ek_2zones}
\end{figure}

In Fig.\,\ref{fig:ek_2zones}, we observe the distribution of kinetic energy in a two-zone interior model. We can see that the internal part of the model follows a gravito-inertial pattern, while the external zone exhibits a single inertial mode. This discrepancy, in comparison to the left panel of Fig.\,\ref{fig:energy},  arises due to the absence of the external stably stratified layers, which theoretically facilitate wave reflection and the formation of two distinct inertial modes. Nevertheless, we can see that the attractor's presence remains consistent, independent of the presence of the narrow intermediate stably stratified region.
Furthermore, we find that the impact of this zone on the dissipation is very weak.

\subsection{Impact of the size of solid and diluted cores}
\begin{figure*}
    \centering
    \resizebox{\hsize}{!}{\includegraphics{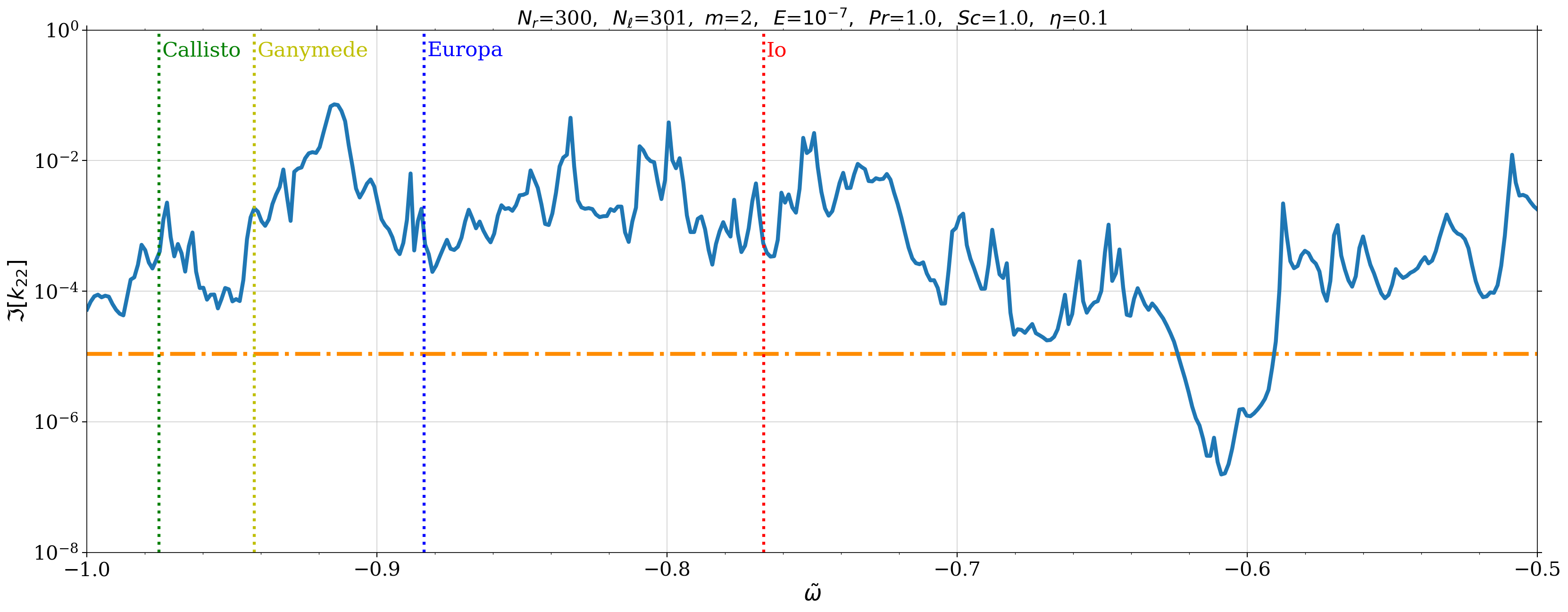}}
    \caption{Same as the middle panel of the Fig.\,\ref{fig:Imk} but for $\eta=0.1$.}
    \label{fig:Imk_eta1}
\end{figure*}

\begin{figure}
    \centering
    \resizebox{\hsize}{!}{\includegraphics{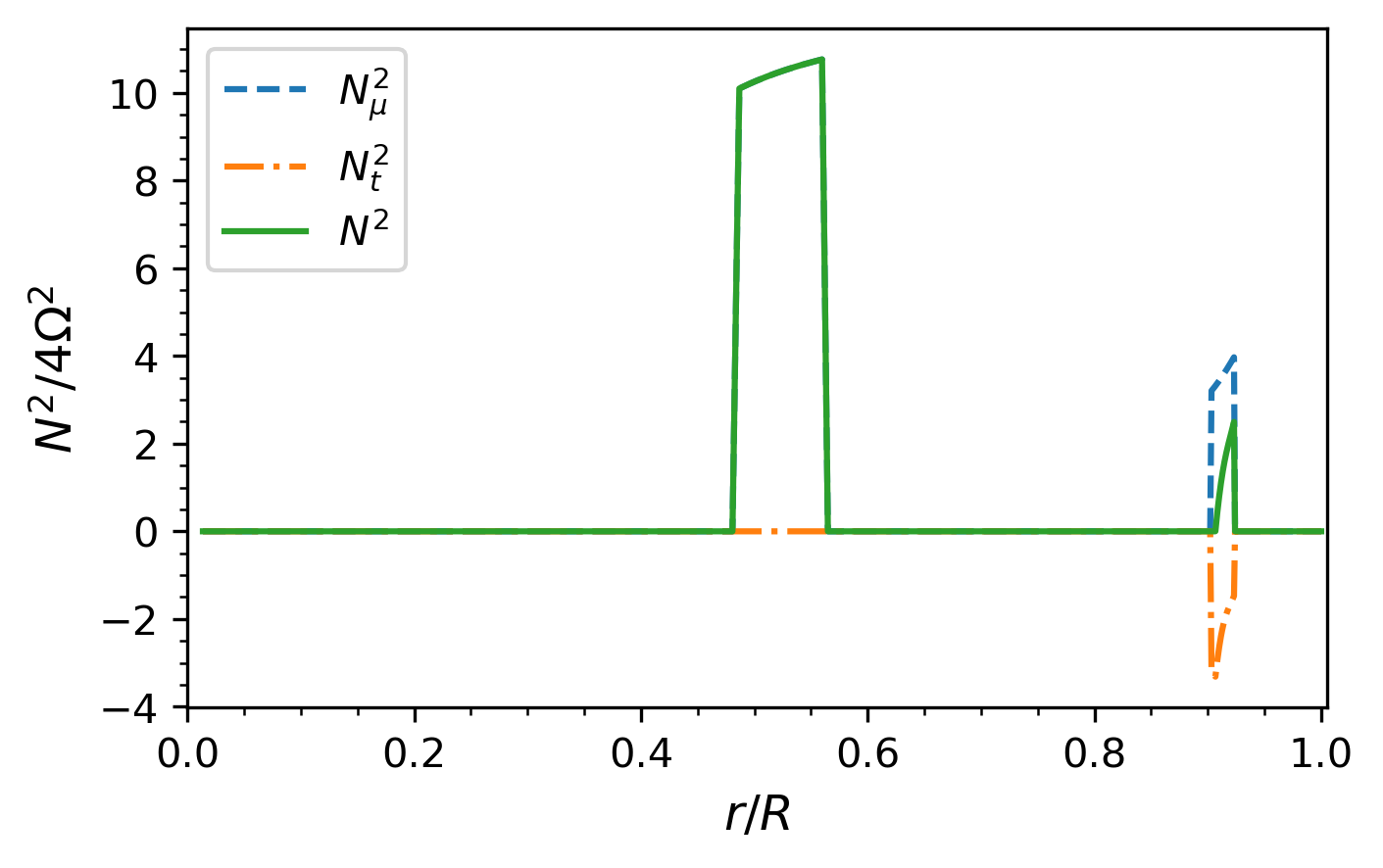}}
    \caption{Same as Fig.\,\ref{fig:freq_N} but for a structure model with a smaller internal stably stratified layer.}
    \label{fig:freq_N_sec_model}
\end{figure}

\begin{figure*}
    \centering
    \resizebox{\hsize}{!}{\includegraphics{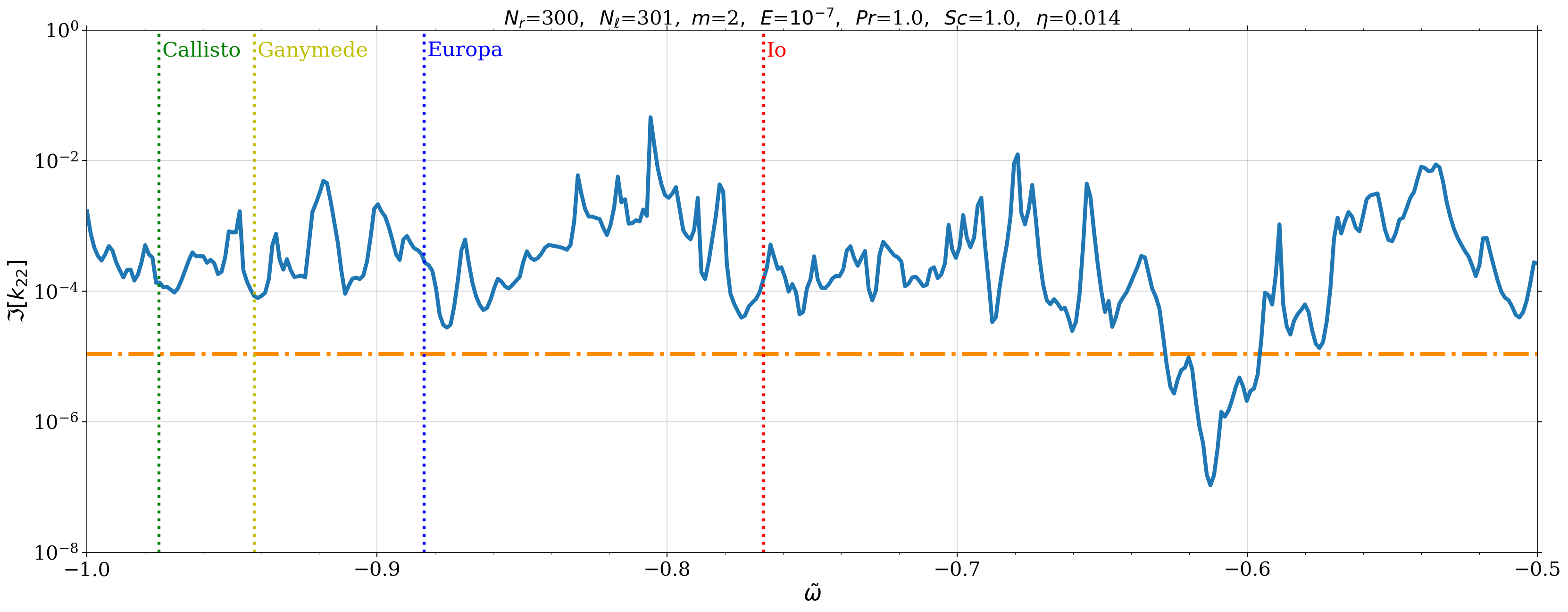}}
    \caption{Same as the middle panel of the Fig.\,\ref{fig:Imk} but for the structure model represented in Fig.\,\ref{fig:freq_N_sec_model}.}
    \label{fig:Imk_cd21}
\end{figure*}

In order to study the impact of the size of the diluted and solid cores on the total dissipation, we first carry out a test with the five-layer model (Sec.\,\ref{subsec:struc_model}), but with a bigger solid core (smaller diluted core) of size 10\% instead of 1.4\%. As we can see in Fig.\,\ref{fig:Imk_eta1}, we find that the magnitude of imaginary part has slightly decreased (the  imaginary part of the Love number due to Io is $5.4 \times 10^{-4}$) and that the position of the peaks is only slightly modified.
We perform another test with the two-layer model, but this time with a solid core of size 15\% instead of 1.4\%. We find that the position of the peaks changes and the dissipation due to Io increases by less than half an order of magnitude (with this model, the imaginary part of the love number due to Io is $2.5 \times 10^{-3}$).
Afterwards, we use another structure model that  satisfies also Juno constraints and uses the equation of state of \cite{Chabrier2021} where we reduce the size of the internal stably stratified layer (diluted core). As we can see in Fig.\,\ref{fig:freq_N_sec_model}, the size of the diluted is reduced, and it is now localised between 48\% and 56\% of the radius (the outer stably stratified layer is in the same position). We can see also that the stratification in this model is stronger ($N^2/(2\Omega)^2 \approx 10$ instead of $\approx 4$). 
With this model, we find, as we can see in Fig.\,\ref{fig:Imk_cd21}, that the imaginary part of the love number due to Io is closer to the observed value (with this model the imaginary part of the Love number due to Io is $1.4 \times 10^{-4}$), but the gap remains significant (approximately one order of magnitude).

\section{Discussion and conclusions}\label{sec:6}

We develop a numerical method that enables the calculation of the forced dynamic tidal response of an incompressible, non-magnetised, uniformly rotating fluid body. The Coriolis force is fully accounted for in our calculations. However, we do not consider centrifugal distortion, which allows us to solve the problem using spherical geometry. We take into consideration various types of dissipations such as fluid viscosity, thermal dissipation, and molecular diffusivity. By incorporating these dissipation mechanisms, we compute, using 2D numerical simulations, the total dissipation and determine the imaginary part of the tidal Love numbers for a given complex planetary interior model.
In this study, we examine the dynamical tides in the latest Jupiter interior model (Sec.\,\ref{sec:3}) and specifically investigate the quadrupolar tidal components ($\ell=m=2$). Our focus is on the frequency range that corresponds to the tidal frequencies associated with Jupiter's Galilean moons.
We consider a multi-layer model with alternating convective and stably stratified regions, which enables a more comprehensive and realistic representation of the physical processes occurring within giant gaseous planets’ interiors, in particular the dissipation of dynamical tides. 
We find that the presence of stably stratified regions plays a significant role in explaining the strong dissipation observed in Jupiter when compared to the case of a sole convective envelope. 
In this framework, we find that the dissipation depends on the chosen internal structure, in particular the size of the diluted core. In fact, with a large diluted core (around 68\% of the radius) we find a two-order-of-magnitude discrepancy between the calculated and observed dissipation due to Io, whereas with a smaller stably stratified inner layer (around 8\% of the radius), the discrepancy becomes smaller (one order of magnitude). This may provide in the future constrains on the size of the diluted core.
Our analysis reveals also that, in the chosen parameter regime in which the kinematic viscosity, thermal and molecular diffusivities are uniform and equal (the realistic variation in transport coefficients vary by several orders of magnitude and their ratios are potentially different from 1 depending on the considered region), the dominant mechanism contributing to dissipation is viscosity, surpassing both thermal and chemical dissipations in magnitude.
Furthermore, it is important to note that our model is not limited to Jupiter but can also be applied to other giant planets such as Saturn, as well as exoplanets.

There are several caveats that should be carefully considered in future studies in order to ensure accurate quantitative comparisons with high-precision observations, it is crucial to incorporate  the relevant missing physical processes in a self-consistent manner.
First, neglecting the influence of centrifugal effects may limit the accuracy of our solutions. Particularly for high-degree tidal components, the impact of centrifugal forces becomes increasingly significant \citep{Dewberry2023}. Second, while adopting the Boussinesq approximation to investigate dynamical tides simplifies the system of equations to solve, it is important to acknowledge its limitations. These limitations are particularly significant when the Lamb frequency, which characterizes the acoustic modes, approaches near the surface a comparable magnitude to the excited modes frequencies. Clearly, an important follow-up of this work would be to go from the Boussinesq approximation to the anelastic approximation and take into account density stratification. The outcomes of using the more realistic anelastic approximation are not expected to completely deviate from those obtained with the Boussinesq approximation; in fact, both approximations yield the same attractors of characteristics. This comes about because in the anelastic approximation the velocity $\Vec{v}$ is replaced by the specific linear momentum $\rho\Vec{v}$ in the system of equations. Said differently, the momentum vector satisfies the same set of equations as the velocity vector does in the Boussinesq case \citep{Dintrans2000}. 
In this respect, using simple polytropic models, the work of \cite{Ogilvie2013} gives a first exploration of the effects of density variations of the background on tidal dissipation. The Boussinesq approximation may overestimate the tidal dissipation that could explain why the computed dissipation in our work is too large when compared to the observations. This will be carefully evaluated in a following work where we shall use the anelastic approximation.

Finally, differential rotation can play a crucial role in the dynamics of the outer regions of gas giant planets. In addition, the ionized inner region, characterized by the presence of a magnetized gas, can exhibit significant effects due to ohmic dissipation and induced magnetic torques. 
We know that the presence of differential rotation in a convective zone is strongly dependent on electrical conductivity \citep[e.g.][]{Guillot2018, Galanti2019}. 
Therefore, an interesting perspective of this work is to undertake a study to understand the profound impact of both differential rotation \citep{Mathis2009, Baruteau2013, Mirouh2016, Guenel2016a, Guenel2016b, Dewberry2021} and magnetic fields \citep{Rogers2010, Mathis2011, Barker2014, Wei2016, Wei2018, Lin2018} on wave propagation and dissipation in gas giant planets.

\begin{acknowledgements}
We thank the referee for her/his positive and constructive report, which has allowed us to improve the quality of our article. H. Dhouib and S. Mathis acknowledge support from the CNES PLATO grant at CEA/DAp and from PNP (CNRS/INSU). A. Astoul acknowledges support from the Science and Technology Facilities Council (STFC) grant ST/S000275/1, as well as the Leverhulme Trust for early career grant. We are also very grateful to Lorenzo Valdettaro for his kind help and support with the LSB code. Some preliminary calculations have been performed thanks to HPC resources from CALMIP supercomputing centre (Grants P16024 and 2022-P0107). S. Mathis and M. Rieutord acknowledge support from the European Research Council through HORIZON ERC SyG Grant 4D-STAR 101071505.
\end{acknowledgements}

\bibliographystyle{aa}
\bibliography{biblio}

\begin{appendix}
\section{Dimensionless number expressions in Jupiter}
\label{append:numb_exp}
The expressions for the different diffusivities and associated dimensionless numbers are derived in \cite{Stevenson1977} for metallic and molecular phases.
The region with $R<0.9$ exhibits a metallic phase, whereas the region with $R>0.9$ is characterized by a molecular phase. We add a smooth transition between both zones. We recall here the expression for these numbers
\begin{equation}
    \mathrm{Pr}= \left\{
    \begin{array}{ll}
        \displaystyle{\frac{4}{0.3}T^{-1/2}\rho^{-1/3}} & \mbox{if } R<0.9 \\ \\
        1 & \mbox{if not}
    \end{array}
\right.,
\end{equation}

\begin{equation}
    \mathrm{Sc}= \left\{
    \begin{array}{ll}
        \displaystyle{\frac{4}{3}10^{6}T^{-2}\rho^{2/3}} & \mbox{if } R<0.9 \\ \\
        \displaystyle{\frac{1}{4}10^{6.5}T^{-2}\rho^{5/6}} & \mbox{if not}
    \end{array}
\right.,
\end{equation}

\begin{equation}
    \mathrm{E}= \left\{
    \begin{array}{ll}
        \displaystyle{2\times10^{-5}\frac{ T^{-1/2}}{\Omega R^2}} & \mbox{if } R<0.9 \\ \\
        \displaystyle{5\times 10^{-5}\frac{T^{-1/2}}{\Omega R^2}} & \mbox{if not}
    \end{array}
\right..
\end{equation}
All quantities must be expressed in SI units.

\section{Expression of some operators in the spherical harmonics basis} \label{append:exp_harmo}
The spherical vector harmonics form a complete family and the orthogonality relations ensure that any sufficiently regular vector field $\Vec{u}$ can be uniquely expanded over the spherical vector harmonics \citep{Rieutord1987} :
\begin{equation}
    \vec{u}=\left[u_\ell^m,\; v_\ell^m,\; w_\ell^m\right].
\end{equation}
The divergence of the vector field $\Vec{u}$
\begin{equation}
    \vec{\nabla} \cdot \vec{u}=\frac{1}{r^2} \partial_r (r^2 u_\ell^m)-\frac{\ell(\ell+1)}{r}v_\ell^m.
\end{equation}
The curl of the vector field $\vec{u}$ is expressed as follows
\begin{equation}
    \vec{\nabla} \times \vec{u}=\left[\ell(\ell+1)\frac{w_\ell^m}{r},\; \frac{1}{r} \partial_r (r w_\ell^m),\; \frac{u_\ell^m}{r}-\frac{1}{r} \partial_r (r v_\ell^m)\right].
\end{equation}

\subsection*{Particular case: $\vec{\nabla} \cdot \vec{u}=0$}
The curl of the vector field $\vec{u}$ can be rewritten as
\begin{equation}
    \vec{\nabla} \times \vec{u} = \left[\ell(\ell+1) \frac{w_\ell^m}{r},\; \frac{1}{r} \partial_r (r w_\ell^m),\; -\frac{\Delta_\ell (r u_\ell^m)}{\ell(\ell+1)}\right],
\end{equation}
with $\displaystyle{\Delta_{\ell}\square=\partial_{r^2}^{2}\square+\frac{2}{r} \partial_{r}\square-\frac{\ell(\ell+1)}{r^{2}}\square}$.\\
The Laplacian of the vector field $\vec{u}$ is expressed as follows
\begin{equation}
    \Vec{\nabla}^2\vec{u} = \left[\frac{1}{r}\Delta_\ell \left(r u_\ell^m\right),\; \frac{1}{r} \partial_r \left(\frac{D_\ell u_\ell^m}{\ell(\ell+1)}\right),\; \Delta_\ell w_\ell^m\right],
\end{equation}
with $D_\ell \square = \partial_r^2 \left(r^2\square\right)- \ell(\ell+1)\square$.\\
The vector product between the unit vector $\vec{e}_z$ and the vector field $\vec{u}$ is given by
\begin{multline}
\vec{e}_z \times \vec{u}=\left[(\ell-1) \alpha_{\ell-1}^{\ell} w_{\ell-1}^m- (\ell+2) \alpha_{\ell+1}^{\ell} w_{\ell+1}^m - i m v_\ell^m, \right.\\
 \frac{\ell-1}{\ell} \alpha_{\ell-1}^{\ell} w_{\ell-1}^m + \frac{\ell+2}{\ell+1} \alpha_{\ell+1}^{\ell} w_{\ell+1}^m - \frac{i m}{\ell(\ell+1)}\left(u_\ell^m+v_\ell^m\right), \\
\left. - \frac{\alpha_{\ell-1}^{\ell}}{\ell^2} r^{\ell-1} \partial_r\left(\frac{u_{\ell-1}^m}{r^{\ell-2}}\right) - \frac{\alpha_{\ell+1}^{\ell}}{(\ell+1)^2} r^{-\ell-2} \partial_r\left(r^{\ell+3} u_{\ell+1}^m\right) - \right.\\ \left. \frac{i m}{\ell(\ell+1)} w_\ell^m\right].
\end{multline}

\end{appendix}

\end{document}